\documentclass[12pt]{article}
\usepackage[margin=0.7in]{geometry}

\usepackage{amsmath}
\usepackage{amsfonts}
\usepackage{amsthm}
\usepackage{array}
\usepackage{multirow}
\usepackage{graphicx}
\usepackage{caption}
\usepackage{subcaption}

\DeclareMathOperator{\diver}{div}
\DeclareMathOperator{\Tr}{Tr}

\newcommand{\totalDiff}[1]{\frac{\mathrm{d}}{\mathrm{d}{#1}}}
\newcommand{\dir}[1]{\partial_{ {#1}}}
\newcommand{\LieAlgebra}[1]{\mathfrak{#1}}
\newcommand{\bvec}[1]{\mathbf{#1}}
\newcommand{\entr}{s}
\newcommand{\temp}{T}
\newcommand{\dens}{\rho}
\newcommand{\press}{p}
\newcommand{\energy}{\epsilon}
\newcommand{\systemEk}[1]{\mathcal{E}_{#1}} 
\newcommand{\stress}{\sigma}
\newcommand{\visc}{\sigma_v}

\newtheorem{theorem}{Theorem}

\title{Symmetry classification of viscid flows on space curves}

\author{Anna Duyunova,\\
	Institute of Control Sciences of RAS,
	Bauman Moscow State Technical University, 
	\\anna.duyunova@yahoo.com\\
	Valentin Lychagin,\\
	Institute of Control Sciences of RAS, University of Troms\o, \\
	valentin.lychagin@uit.no,\\
	Sergey Tychkov,
	\\Institute of Control Sciences of RAS,\\sergey.lab06@ya.ru}

\begin{document}
\maketitle
\begin{abstract}
	Symmetries and differential invariants of viscid flows with viscosity
	depending on temperature on a space curve are given. Their dependence on
	thermodynamic states of media is studied, and a classification of
	thermodynamic states is given.
\end{abstract}


\section{Introduction}

In this paper, we continue studies of viscid flows on space curves started in
\cite{DLTcurve}, \cite{DLTcurveNS}. Here, we consider media with viscosity
$\zeta$ being a function of temperature $\temp$. The main goal of the paper is
to provide classification from the admitted symmetries standpoint.

Recall that motion of viscid flows on an oriented Riemannian manifold $(M, g)$
in the field of constant gravitational force satisfies the Navier--Stokes
equations (see \cite{Batchelor2000}, \cite{DLTNS} for details):
\begin{equation}\label{eq:E}
\left\lbrace
\begin{aligned}
&\dens(\bvec{u}_t + \nabla_{\bvec{u}}\bvec{u})-
\diver\stress - \bvec{g}\dens = 0,\\
&\dens_t + \bvec{u}(\dens) + \dens\,\diver\bvec{u} = 0,\\
&\dens\temp(\entr_t + \bvec{u}(s)) -
\Tr (\visc^*(d_{\nabla}\bvec{u})) +
k\,\Delta_g \temp = 0,
\end{aligned}
\right.
\end{equation}
where $\bvec{u}$ is the flow velocity, $\press$, $\dens$, $\entr$, $\temp$,
$\stress=-p\delta_{ij}+\visc$ are the pressure, density, specific entropy,
temperature, stress tensor of the fluid respectively, $k$ is a constant thermal
conductivity and $\bvec{g}$ is the gravitational acceleration field.

We consider a flow on a naturally-parameterized curve 
\[
M = \{x=f(a),\,  y=g(a), \, z=h(a)\} 
\]
in the three-dimensional Euclidean space. In this case vector  $\bvec{g}$ is the
restriction of the vector field $(0,0,\mathrm{g})$ on $M$, i.e. 
\[
\bvec{g} = \mathrm{g}  h^{\prime} \partial_a, \quad
\mathrm{g}<0.
\]

Because the system~\eqref{eq:E} is underdetermined, we use methods described in
\cite{DLTwisla} to get two additional relations between thermodynamic
quantities. The idea of this method is based on interpretation of media
thermodynamic states as Legendrian, or Lagrangian, manifolds in contact, or
symplectic, space correspondingly.


By the Navier--Stokes system $\systemEk{}$, we mean the system \eqref{eq:E}
restricted to the curve $M$ along with Lagrangian manifold $L$ in a
four-dimensional symplectic space endowed with structure form
\[
\Omega = d\entr \wedge d\temp + {\dens^{-2}} d\dens \wedge d\press.
\]

Also we require \cite{LychaginWisla} the restriction $\kappa\vert_L$ of the
quadratic form
\[
\kappa = d(\temp^{-1}) \cdot d\energy -
\dens^{-2} d(\press \temp^{-1}) \cdot d\dens
\]
be negative definite, where $\energy$ is the specific internal energy.

The paper is organized as follows.

In Section \ref{sec:symmetries} we consider two Lie algebras, namely, a symmetry
algebra $\LieAlgebra{g}$ of the system \eqref{eq:E} restricted to the curve $M$
and a symmetry algebra $\LieAlgebra{g_{sym}}$ of the system $\systemEk{}$. Then
we give a classification of symmetry algebras depending on the functions $\zeta$
and $h$. It turns out that, from the admissible symmetry algebra standpoint,
there are three different models of viscosity as a function of temperature and
seven types of curves including arbitrary ones.

The following functions $\zeta$ have distinct symmetry algebras: (a)
$\zeta(\temp) = \alpha \temp$, (b) $\zeta(\temp) = \alpha \temp^{\beta}$,
$\beta\neq 1$, and (c) all others. It is worth to note that two well-known
models of viscosity, namely, the elastic hard-ball model and power-law model
fit into the case (b), which possesses more symmetries. While other models,
including exponential law, are less symmetrical.

Though the case of constant viscosity may be considered as a part of case (b)
given $\beta=0$, sometimes it needs special consideration and additional
computations concerning differential invariants and thermodynamic states, which
can be found in \cite{DLTcurveNS}.

Since the system $\systemEk{}$ and, therefore, its symmetry algebra depend only
on the $z$-component of the curve, that is the function $h$, we may give
geometrical interpretation for the seven special cases of $h$. If a space curve
is represented as a pair of plane curve $(x(\tau), y(\tau))$ and a function
$z(\tau)$ considered as a way of lifting the plain curve, we can extend our
classification of different forms of $h$ to the `lifting' function $z$. It
appears that the distinct cases of $h$ in the present classification match ones
found in \cite{DLTcurve} (though corresponding algebras are different),
therefore, we only recollect connection between the functions $h(a)$ and
$z(\tau)$ in Appendix.

In Section \ref{sec:therm} we consider thermodynamic states that admit a
one-dimensional symmetry algebra. For each case studied in Section
\ref{sec:symmetries} corresponding thermodynamic states are found in the form
of two relations on $\press$, $\dens$, $\temp$ and $\entr$.

In Section \ref{sec:invs} we recall the notion of a differential invariant of
the Navier--Stokes system and introduce two classes of differential invariants,
namely, kinematic and Navier--Stokes invariants. Fields of differential
invariants corresponding to different cases of the function $h$ and $\zeta$ are
described.

All computations for this paper are made with the DifferentialGeometry package
\cite{Anderson2016} in Maple (see the Maple files \textit{http://d-omega.org}).

\section{Symmetry Lie algebra}\label{sec:symmetries}


Evidently, the symmetry Lie algebra $\LieAlgebra{g_{sym}}$
of the system $\systemEk{}$ depends on the functions $\zeta$
and $h$. To describe this Lie algebra, we introduce a Lie
algebra $\LieAlgebra{g}$, which is the algebra of point
symmetries of the PDE system \eqref{eq:E}.

Depending on how a symmetry acts on thermodynamic quantities
we distinct two kinds of point symmetries of the PDE system
$\systemEk{}$. Namely, if a symmetry acts on thermodynamic
phase space trivially, we call it a {\it geometric
symmetry}. The second kind may be specified as follows.
Consider a Lie algebras homomorphism $\vartheta \colon
\LieAlgebra{g} \rightarrow \LieAlgebra{h}$,
\[
\vartheta\colon X\mapsto
X(\dens)\dir{\dens} + X(\entr)\dir{\entr} +
X(\press)\dir{\press} + X(\temp)\dir{\temp},
\]
where $\LieAlgebra{h}$ is a Lie algebra of vector fields on the thermodynamic space  $(\press, \dens, \entr, \temp)$.

The algebra of geometric symmetries $\LieAlgebra{g_{m}}$
coincides with $\ker \vartheta$.

Let $\LieAlgebra{h_{t}}$ be the Lie subalgebra of the
algebra $\LieAlgebra{h}$ that preserves thermodynamic state $L$.

\begin{theorem}[\cite{DLTwisla}]
	The Lie algebra $\LieAlgebra{g_{sym}}$ of symmetries of
	the Navier--Stokes system $\systemEk{}$ coincides with 
	\[
	\vartheta^{-1}(\LieAlgebra{h_{t}}).
	\]
\end{theorem}

\subsection{$\zeta(\temp)$ is an arbitrary function}

First of all, we consider the case of an arbitrary function
$h(a)$. Then the Lie algebra $\LieAlgebra{g}$ is generated
by the vector fields 
\begin{equation*} 
X_1 = \dir{ t}, \qquad  
X_2 = \dir{ \press} , \qquad
X_3 = \dir{ \entr}  .
\end{equation*}

The corresponding pure thermodynamic part $\LieAlgebra{h}$
of the symmetry algebra is generated by
\begin{equation*} 
Y_1 = \dir{\press}, \qquad Y_2 = \dir{\entr}.
\end{equation*}

We may conclude that the system $\systemEk{}$ admits the
smallest Lie algebra of point symmetries $\vartheta^{-1}
(\LieAlgebra{h_{t}})$, when the function $h(a)$ is
arbitrary.

It is natural to expect that the algebra $\vartheta^{-1}
(\LieAlgebra{h_{t}})$ will be larger for some special forms
of $h$. These special cases are listed below.

\medskip
\textbf{1.} $h(a) = const$.

The Lie algebra $\LieAlgebra{g}$ is generated by 
\[
X_1,\quad
X_2,\quad
X_3,\quad
X_4 = \dir{ a}, \quad
X_5 = t\,\dir{ a}+\dir{ u}, \quad
X_6 = t\,\dir{ t}+a\,\dir{a}-\press\,\dir{ \press} -
\dens\,\dir{ \dens}.
\]

This Lie algebra is solvable and its sequence of derived algebras is
\[
\LieAlgebra{g} = \left\langle X_1,X_2,\ldots,X_6 \right\rangle \supset
\left\langle X_1,X_2,X_4\right\rangle\supset 0.
\]

The pure thermodynamic part $\LieAlgebra{h}$ is generated by
\[
Y_1 = \dir{\press}, \quad 
Y_2 = \dir{\entr}, \quad
Y_3 = \press\,\dir{ \press} + \dens\,\dir{ \dens}.
\]

\textbf{2.} $h(a)= \lambda a$,  $\lambda\neq 0$

The Lie algebra $\LieAlgebra{g}$ is generated by
\[
X_1,\quad
X_2,\quad
X_3,\quad
X_4 = \dir{ a}, \quad
X_5 = t\,\dir{ a}+\dir{ u}, \quad 
X_6 = t\,\dir{ t}+ \left(\frac{\lambda g t^2}{2}+a\right)\,\dir{ a}
+\lambda g t\,\dir{ u} -\press\,\dir{ \press} - \dens\,\dir{ \dens}.
\]

This Lie algebra is solvable and its sequence of derived algebras is
\[
\LieAlgebra{g} = \left\langle X_1,X_2,\ldots,X_6 \right\rangle \supset
\left\langle X_2,X_4,X_1+\lambda g X_5 \right\rangle  \supset0.
\]

The pure thermodynamic part $\LieAlgebra{h}$ is generated by
\begin{equation*} 
Y_1 = \dir{ \press}, \qquad  
Y_{2} = \dir{ \entr} , \qquad  
Y_{3} = \press\,\dir{ \press} + \dens\,\dir{ \dens} .
\end{equation*}

\textbf{3.} $h(a)= \lambda a^2$, $\lambda\neq 0$

If $\lambda<0$, the Lie algebra $\LieAlgebra{g}$ is generated by 
\[
\begin{aligned} 
&X_1,\quad
X_2,\quad
X_3,\quad\\
&X_4 = \sin(\sqrt{2\lambda \mathrm{g}}\,t)\,\dir{a} + 
\sqrt{2\lambda \mathrm{g}} \cos(\sqrt{2\lambda \mathrm{g}}\,t)\,\dir{ u} , \quad 
X_5 =\cos(\sqrt{2\lambda \mathrm{g}}\,t)\,\dir{a} -
\sqrt{2\lambda \mathrm{g}} \sin(\sqrt{2\lambda \mathrm{g}}\,t)\,\dir{ u} , 
\end{aligned} 
\]
and, if $\lambda>0$,  by
\[
\begin{aligned} 
&X_1,\quad
X_2,\quad
X_3,\quad\\
&X_4 = e^{\sqrt{-2\lambda \mathrm{g}}\,t}\dir{a} +
\sqrt{-2\lambda \mathrm{g}}\, e^{\sqrt{-2\lambda \mathrm{g}}\,t}\dir{ u} , \quad 
X_5 =e^{-\sqrt{-2\lambda \mathrm{g}}\,t}\dir{a} -
\sqrt{-2\lambda \mathrm{g}} \, e^{-\sqrt{-2\lambda \mathrm{g}}\,t}\dir{ u} . 
\end{aligned}  
\]

This Lie algebra is solvable and its sequence of derived algebras is
\[
\LieAlgebra{g} = \left\langle X_1,X_2,\ldots,X_5 \right\rangle\supset
\left\langle X_4,X_5 \right\rangle \supset0.
\]

The pure thermodynamic part $\LieAlgebra{h}$ is generated by
\begin{equation*} 
Y_1 = \dir{ \press}, \qquad  
Y_{2} = \dir{ \entr} .
\end{equation*}

\textbf{4.} $h(a)= \ln a$

The Lie algebra $\LieAlgebra{g}$ is generated by
\[
X_1,\quad
X_2,\quad
X_3,\quad
X_4  = t\,\dir{ t}+ a  \,\dir{a}- \press\,\dir{ \press} -\dens\,\dir{\dens}.
\]

This Lie algebra is solvable and its sequence of derived algebras is
\[
\LieAlgebra{g} = \left\langle X_1,X_2,X_3,X_4 \right\rangle \supset
\left\langle X_1,X_2 \right\rangle \supset0.
\]

The pure thermodynamic part $\LieAlgebra{h}$ is generated by
\begin{equation*} 
Y_1 = \dir{ \press}, \qquad  
Y_{2} = \dir{ \entr} , \qquad  
Y_{3} =  \press\,\dir{ \press} +\dens\,\dir{\dens} .
\end{equation*}

The table below sums up the results of this subsection.

\vspace*{12pt}
\begingroup
\setlength{\tabcolsep}{2pt}
\renewcommand{\arraystretch}{2.9}
\begin{tabular}{||b{0.24\linewidth}|b{0.6\linewidth}||}
	\hline
	$h(a)$ is arbitrary
	&$\begin{aligned} 
	 \dir{ t}, \quad
	 \dir{\press }, \quad 
	 \dir{\entr}
	\end{aligned} $ \\
	\hline
	$h(a)= const$ 
	&$\begin{aligned} 
	& \dir{ t}, \quad
	\dir{\press }, \quad 
	\dir{\entr}, \quad \dir{ a}, \quad
	 t\,\dir{ a}+\dir{ u}, \quad 
	 t\,\dir{ t}+a\,\dir{a}-\press\,\dir{ \press} - \dens\,\dir{ \dens} 
	\end{aligned} $ \\
	\hline
	$h(a)= \lambda a$, $\lambda\neq 0 $ 
	& $\begin{aligned} 
	&\dir{ t}, \quad
	\dir{\press }, \quad 
	\dir{\entr}, \quad \dir{ a}, \quad
	 t\,\dir{ a}+\dir{ u}, \quad \\
	& t\,\dir{ t}+ \left(\frac{\lambda g t^2}{2}+a\right)\,\dir{ a} +
	\lambda g t\,\dir{ u} -\press\,\dir{ \press} - \dens\,\dir{ \dens}
	\end{aligned} $ \\
	\hline
	$h(a)= \lambda a^2$, $\lambda\neq 0 $
	&$\begin{aligned} 
	&\dir{ t}, \quad
	\dir{\press }, \quad 
	\dir{\entr}, \quad \\
	&e^{\sqrt{2\lambda g}\,t}\dir{a} +  \sqrt{2\lambda g} \,
	e^{\sqrt{2\lambda g}\,t}\dir{ u} , \quad 
	e^{-\sqrt{2\lambda g}\,t}\dir{a} -  \sqrt{2\lambda g} \, 
	^{-\sqrt{2\lambda g}\,t}\dir{ u}  
	\end{aligned} $\\
	\hline
	$h(a)= \ln{a}$ 
	&$\begin{aligned} 
	&\dir{ t}, \quad
	\dir{\press }, \quad 
	\dir{\entr}, \quad
	 t\,\dir{ t}+ a  \,\dir{a}- \press\,\dir{ \press} -\dens\,\dir{\dens}
	\end{aligned}$\\
	\hline
\end{tabular}
\vspace{12pt}
\endgroup

\subsection{$\zeta(\temp)= \alpha \temp$} 

If the medium viscosity is proportional to the temperature,
the symmetry algebra
differs from the previous case only
in one additional symmetry $\press\,\dir{ \press} +
\dens\,\dir{ \dens} - \entr\,\dir{ \entr} + \temp\,\dir{ \temp}$.
See table below.

\vspace*{12pt}
\begingroup
\setlength{\tabcolsep}{2pt}
\renewcommand{\arraystretch}{2.9}
\begin{tabular}{||b{0.24\linewidth}|b{0.6\linewidth}||}
	\hline
	$h(a)$ is arbitrary
	&$\begin{aligned} 
	&\dir{ t}, \quad
	 \dir{\press }, \quad 
	 \dir{\entr}, \quad
	 \press\,\dir{ \press} + \dens\,\dir{ \dens} - \entr\,\dir{ \entr} + \temp\,\dir{ \temp}
	\end{aligned} $ \\
	\hline
	$h(a)= const$ 
	&$\begin{aligned} 
	&\dir{ t}, \quad
	\dir{\press }, \quad 
	\dir{\entr}, \quad
	\dir{ a}, \quad
	t\,\dir{ a}+\dir{ u},\\
	&\press\,\dir{ \press} + \dens\,\dir{ \dens} - \entr\,\dir{ \entr} + \temp\,\dir{ \temp}, \quad 
	 t\,\dir{ t}+a\,\dir{a}-\press\,\dir{ \press} - \dens\,\dir{ \dens} 
	\end{aligned} $ \\
	\hline
	$h(a)= \lambda a$, $\lambda\neq 0 $ 
	& $\begin{aligned} 
	&\dir{ t}, \quad \dir{ a}, \quad
	\dir{\press }, \quad 
	\dir{\entr}, \quad
	\press\,\dir{ \press} + \dens\,\dir{ \dens} - \entr\,\dir{ \entr} + \temp\,\dir{ \temp}, \\
	& t\,\dir{ a}+\dir{ u}, \quad
	t\,\dir{ t}+ \left(\frac{\lambda g t^2}{2}+a\right)\,\dir{ a} +\lambda g t\,\dir{ u} -\press\,\dir{ \press} - \dens\,\dir{ \dens}
	\end{aligned} $ \\
	\hline
	$h(a)= \lambda a^2$, $\lambda\neq 0 $
	&$\begin{aligned} 
	&\dir{ t}, \quad
	\dir{\press }, \quad 
	\dir{\entr},  \quad
	\press\,\dir{ \press} + \dens\,\dir{ \dens} - \entr\,\dir{ \entr} + \temp\,\dir{ \temp}, \\
	&e^{\sqrt{2\lambda g}\,t}\dir{a} +  \sqrt{2\lambda g} \, e^{\sqrt{2\lambda g}\,t}\dir{ u} , \quad 
	e^{-\sqrt{2\lambda g}\,t}\dir{a} -  \sqrt{2\lambda g} \, e^{-\sqrt{2\lambda g}\,t}\dir{ u}  
	\end{aligned} $\\
	\hline
	$h(a)= \ln{a}$ 
	&$\begin{aligned} 
	&\dir{ t}, \quad
	\dir{\press }, \quad 
	\dir{\entr}, \quad
	\press\,\dir{ \press} + \dens\,\dir{ \dens} - \entr\,\dir{ \entr} + \temp\,\dir{ \temp},\\
	& t\,\dir{ t}+ a  \,\dir{a}- \press\,\dir{ \press} -\dens\,\dir{\dens}
	\end{aligned}$\\
	\hline
\end{tabular}
\vspace{12pt}
\endgroup

\subsection{$\zeta(\temp)= \alpha \temp^{\beta}, \, \beta\neq 1$}
First of all, we consider the case of an arbitrary function
$h$. Then the Lie algebra $\LieAlgebra{g}$ is generated
by the vector fields 
\begin{equation*} 
X_1 = \dir{ t}, \qquad  
X_{2} = \dir{ \press} , \qquad  
X_{3} =  \dir{ \entr}  .
\end{equation*}

The corresponding pure thermodynamic part $\LieAlgebra{h}$
of the symmetry algebra is generated by
\begin{equation*} 
Y_1 = \dir{ \press}, \qquad  
Y_{2} = \dir{ \entr}  .
\end{equation*}

As before, we see that an arbitrary function $h$ corresponds to the
smallest Lie algebra of point symmetries $\vartheta^{-1}
(\LieAlgebra{h_{t}})$.

Below, the special cases of the function $h$ are listed.

\medskip
\textbf{1.} $h(a) = const$

The Lie algebra $\LieAlgebra{g}$ is generated by
\[
\begin{aligned} 
&X_1, \quad X_2, \quad X_3, \quad
X_4 = \dir{ a}, \quad
X_5 = t\,\dir{ a}+\dir{ u}, \quad
X_6 = t\,\dir{ t}+a\,\dir{a}-\press\,\dir{\press} -\dens\,\dir{ \dens}, \\
&
X_{7} = a\,\dir{ a} + u\,\dir{ u} 
-\frac{2\beta}{\beta-1} \press\,\dir{ \press}  -
\frac{4\beta-2  }{\beta-1}\dens\,\dir{ \dens} +
\frac{2\beta}{\beta-1} \entr\,\dir{ \entr} -
\frac{2  }{\beta-1}\temp\,\dir{ \temp}.
\end{aligned} 
\]

This Lie algebra is solvable and its sequence of derived algebras is
\[
\LieAlgebra{g} = \left\langle X_1,X_2,\ldots,X_7 \right\rangle \supset
\left\langle X_1,X_2,X_3,X_4,X_5 \right\rangle \supset
\left\langle X_4\right\rangle  \supset0.
\]

The pure thermodynamic part $\LieAlgebra{h}$ is generated by
\begin{equation*} 
Y_1 = \dir{ \press}, \qquad  
Y_{2} = \dir{ \entr} , \qquad  
Y_{3} = \press\,\dir{ \press}+\dens\,\dir{ \dens} ,\qquad 
Y_{4} = (\beta-1)\dens\,\dir{ \dens} - \beta \entr\,\dir{ \entr} + \temp\,\dir{ \temp} .
\end{equation*}

\textbf{2.} {$h(a)= \lambda a$,  $\lambda\neq 0$}

The Lie algebra $\LieAlgebra{g}$ is generated by
\[
\begin{aligned} 
&X_1, \quad X_2, \quad X_3, \quad
X_4 = \dir{ a}, \quad
X_5 = t\,\dir{ a}+\dir{ u}, \qquad \\
&X_6 = t\,\dir{ t}+2a\,\dir{a} +  u\,\dir{ u}  -\frac{3\beta-1}{\beta-1} \press\,\dir{ \press}  -
\frac{5\beta-3 }{\beta-1}\dens\,\dir{ \dens} +
\frac{2\beta}{\beta-1} \entr\,\dir{ \entr} -
\frac{2  }{\beta-1}\temp\dir{ \temp} , \\
&X_{7} =  t\,\dir{t}+ \left(\frac{\lambda g t^2}{2}+a\right)\dir{ a} + \lambda g t\,\dir{ u}-\press\,\dir{ \press} - \dens\,\dir{ \dens} .
\end{aligned} 
\]

This Lie algebra is solvable and its sequence of derived algebras is
\[
\LieAlgebra{g} = \left\langle X_1,X_2,\ldots,X_7 \right\rangle \supset
\left\langle X_1,X_2,X_3,X_4,X_5 \right\rangle \supset
\left\langle X_4\right\rangle \supset0.
\]

The pure thermodynamic part $\LieAlgebra{h}$ is generated by
\begin{equation*} 
Y_1 = \dir{ \press}, \qquad  
Y_{2} = \dir{ \entr} , \qquad  
Y_{3} = \press\,\dir{ \press}+\dens\,\dir{ \dens} ,\qquad 
Y_{4} = (\beta-1)\dens\,\dir{ \dens} - \beta \entr\,\dir{ \entr} + \temp\,\dir{ \temp} .
\end{equation*}

\textbf{3.} $h(a)= \lambda a^2$, $\lambda\neq 0$

If $\lambda<0$, the Lie algebra $\LieAlgebra{g}$ is generated by
\begin{align*}
&X_1, \quad X_2, \quad X_3, \quad \\
&X_4 = \sin(\sqrt{2\lambda \mathrm{g}}\,t)\,\dir{a} +  \sqrt{2\lambda \mathrm{g}} \cos(\sqrt{2\lambda \mathrm{g}}\,t)\,\dir{ u} , \quad 
X_5 =\cos(\sqrt{2\lambda \mathrm{g}}\,t)\,\dir{a} -  \sqrt{2\lambda \mathrm{g}} \sin(\sqrt{2\lambda \mathrm{g}}\,t)\,\dir{ u} , \\
&X_6 = a\, \dir{ a} +u\,\dir{ u}  
-\frac{2\beta}{\beta-1} \press\,\dir{ \press}  -
\frac{4\beta-2  }{\beta-1}\dens\,\dir{ \dens} +
\frac{2\beta}{\beta-1} \entr\,\dir{ \entr} -
\frac{2  }{\beta-1}\temp\dir{ \temp} 
\end{align*} 
and, if $\lambda>0$,  by 
\begin{align*} 
&X_1, \quad X_2, \quad, X_3, \quad \\
&X_4 = e^{\sqrt{-2\lambda \mathrm{g}}\,t}\dir{a} +  \sqrt{-2\lambda \mathrm{g}} \,e^{\sqrt{-2\lambda \mathrm{g}}\,t}\dir{ u} , \quad 
X_5 =e^{-\sqrt{-2\lambda \mathrm{g}}\,t}\dir{a} -  \sqrt{-2\lambda \mathrm{g}} \,e^{-\sqrt{-2\lambda \mathrm{g}}\,t}\dir{ u} , \\
&X_6 = a\, \dir{ a} +u\,\dir{ u}  
-\frac{2\beta}{\beta-1} \press\,\dir{ \press}  -
\frac{4\beta-2  }{\beta-1}\dens\,\dir{ \dens} +
\frac{2\beta}{\beta-1} \entr\,\dir{ \entr} -
\frac{2  }{\beta-1}\temp\dir{ \temp} .
\end{align*}

This Lie algebra is solvable and its sequence of derived algebras is
\[
\LieAlgebra{g} = \left\langle X_1,X_2,\ldots,X_6 \right\rangle\supset
\left\langle X_2,X_3,X_4,X_5 \right\rangle \supset0.
\]

The pure thermodynamic part $\LieAlgebra{h}$ is generated by
\begin{equation*} 
Y_1 = \dir{ \press}, \qquad  
Y_{2} = \dir{ \entr} , \qquad  
Y_{3} = \beta\press\,\dir{ \press}+ (2\beta-1)\dens\,\dir{ \dens} - \beta \entr\,\dir{ \entr} + \temp\,\dir{ \temp} .
\end{equation*}

\textbf{4.} $h(a)= \lambda_1a^{\lambda_2}$, $\lambda_2\neq 0,1,2$ 

The Lie algebra $\LieAlgebra{g}$ is generated by
\[
\begin{aligned}
&X_1, \quad X_2, \quad X_3, \quad \\
&X_4 = t\,\dir{ t} -\frac{2a}{\lambda_2-2} \dir{ a} - \frac{\lambda_2 u }{\lambda_2-2}\dir{ u} + 
\frac{\lambda_2(\beta+1) + 2(\beta-1) }{(\beta-1)(\lambda_2-2)} \press\,\dir{ \press}  + \\
&\phantom{kjhjkhjkhkjjjjhk}
\frac{\lambda_2(3\beta-1) + 2(\beta-1)  }{(\beta-1)(\lambda_2-2)}\dens\,\dir{ \dens} -
\frac{2\beta\lambda_2  }{(\beta-1)(\lambda_2-2)}\entr\,\dir{ \entr}
+ \frac{2\lambda_2  }{(\beta-1)(\lambda_2-2)}\temp\,\dir{ \temp} .
\end{aligned}
\]

This Lie algebra is solvable and its sequence of derived algebras is
\[
\LieAlgebra{g} = \left\langle X_1,X_2,X_3,X_4 \right\rangle\supset
\left\langle X_1,X_2,X_3 \right\rangle\supset0 .
\]

The pure thermodynamic part $\LieAlgebra{h}$ is generated by
\begin{align*} 
&Y_1 = \dir{ \press}, \qquad  
Y_{2} = \dir{ \entr} , \\
&Y_{3} = \left( \lambda_2(\beta+1) + 2(\beta-1)\right) \press\,\dir{ \press}+ (\lambda_2(3\beta-1) + 2(\beta-1))\dens\,\dir{ \dens} - 2\beta\lambda_2 \entr\,\dir{ \entr} + 2\lambda_2 \temp\,\dir{ \temp} .
\end{align*}

\textbf{5.} $h(a)=\lambda_1e^{\lambda_2a}, \, \lambda_2\neq 0$

The Lie algebra $\LieAlgebra{g}$ is generated by
\begin{align*}
&X_1, \quad X_2, \quad X_3, \quad \\
&X_4 = t\,\dir{ t}-\frac{2	}{\lambda_2}\,\dir{a} - u\,\dir{ u}  +\frac{\beta+1}{\beta-1} \press\,\dir{ \press}  +
\frac{3\beta-1 }{\beta-1}\dens\,\dir{ \dens} -
\frac{2\beta}{\beta-1} \entr\,\dir{ \entr} +
\frac{2  }{\beta-1}\temp\dir{ \temp}.
\end{align*}

This Lie algebra is solvable and its sequence of derived algebras is
\[
\LieAlgebra{g} = \left\langle X_1,X_2,X_3,X_4 \right\rangle\supset
\left\langle X_1,X_2,X_3 \right\rangle\supset0 .
\]

The pure thermodynamic part $\LieAlgebra{h}$ is generated by
\[
Y_1 = \dir{ \press}, \qquad  
Y_{2} = \dir{ \entr} , \qquad
Y_{3} =(\beta+1) \press\,\dir{ \press}+ (3\beta-1) \dens\,\dir{ \dens} - 2\beta \entr\,\dir{ \entr} +2 \temp\,\dir{ \temp} .
\]

\textbf{6.} $h(a)= \ln a$

The Lie algebra $\LieAlgebra{g}$ is generated by
\[
X_1, \quad X_2, \quad X_3, \quad
X_4= t\,\dir{ t}+ a  \,\dir{a}- \press\,\dir{ \press} -\dens\,\dir{\dens}.
\]

This Lie algebra is solvable and its sequence of derived algebras is
\[
\LieAlgebra{g} = \left\langle X_1,X_2,X_3,X_4 \right\rangle \supset
\left\langle X_1,X_2 \right\rangle \supset0.
\]

The pure thermodynamic part $\LieAlgebra{h}$ is generated by
\begin{equation*} 
Y_1 = \dir{ \press}, \qquad  
Y_{2} = \dir{ \entr} , \qquad  
Y_{3} = \press\,\dir{ \press}+\dens\,\dir{ \dens}.
\end{equation*}

The following table summarizes the results of this subsection.

\vspace*{12pt}
\begingroup
\setlength{\tabcolsep}{2pt}
\renewcommand{\arraystretch}{2.9}
\begin{tabular}{||b{0.23\linewidth}|b{0.68\linewidth}||}
	\hline
	$h(a)$ is arbitrary
	&$\begin{aligned} 
	&\dir{ t}, \quad
	 \dir{\press }, \quad 
	 \dir{\entr}
	\end{aligned} $ \\
	\hline
	$h(a)= const$ 
	&$\begin{aligned} 
	&\dir{ t}, \quad
	\dir{\press }, \quad 
	\dir{\entr},\quad
	 \dir{ a}, \quad
     t\,\dir{ a}+\dir{ u}, \quad 
	 t\,\dir{ t}+a\,\dir{a}-\press\,\dir{ \press} - \dens\,\dir{ \dens}, \\
	&a\,\dir{ a} + u\,\dir{ u} 
	-\frac{2\beta}{\beta-1} \press\,\dir{ \press}  -
	\frac{4\beta-2  }{\beta-1}\dens\,\dir{ \dens} +
	\frac{2\beta}{\beta-1} \entr\,\dir{ \entr} -
	\frac{2  }{\beta-1}\temp\dir{ \temp} 
	\end{aligned} $ \\
	\hline
	$h(a)= \lambda a$, $\lambda\neq 0 $ 
	& $\begin{aligned} 
	&\dir{ t}, \quad
	\dir{\press }, \quad 
	\dir{\entr},\quad
    \dir{ a}, \quad
    t\,\dir{ a}+\dir{ u},  \\
	&t\,\dir{ t}+2a\,\dir{a} +  u\,\dir{ u}  -
	\frac{3\beta-1}{\beta-1} \press\,\dir{ \press}  -
	\frac{5\beta-3 }{\beta-1}\dens\,\dir{ \dens} + 
	\frac{2\beta \entr}{\beta-1} \dir{ \entr} -
	\frac{2 \temp }{\beta-1}\dir{ \temp} , \\
	& t\,\dir{ t}+ \left(\frac{\lambda g t^2}{2}+a\right)\,\dir{ a} +\lambda g t\,\dir{ u} 
	-\press\,\dir{ \press} - \dens\,\dir{ \dens} 
	\end{aligned} $ \\
	\hline
	$h(a)= \lambda a^2$, $\lambda\neq 0 $
	&$\begin{aligned} 
	&\dir{ t}, \quad
	\dir{\press }, \quad 
	\dir{\entr}, \quad \\
	&e^{\sqrt{2\lambda g}\,t}\dir{a} +  \sqrt{2\lambda g} \, e^{\sqrt{2\lambda g}\,t}\dir{ u} , \quad 
	e^{-\sqrt{2\lambda g}\,t}\dir{a} -  \sqrt{2\lambda g} \, e^{-\sqrt{2\lambda g}\,t}\dir{ u},  \\
	&a\, \dir{ a} +u\,\dir{ u}  
	-\frac{2\beta}{\beta-1} \press\,\dir{ \press}  -
	\frac{4\beta-2  }{\beta-1}\dens\,\dir{ \dens} +
	\frac{2\beta}{\beta-1} \entr\,\dir{ \entr} -
	\frac{2  }{\beta-1}\temp\dir{ \temp} 
	\end{aligned} $\\
	\hline
	$\begin{aligned} 
	&h(a)= \lambda_1a^{\lambda_2},\\ 
	&\lambda_2\neq 0,1,2 
	\end{aligned}$
	&$\begin{aligned} 
	&\dir{ t}, \quad
	\dir{\press }, \quad 
	\dir{\entr}, \quad \\
	&t\,\dir{ t} -\frac{2a}{\lambda_2-2} \dir{ a} - \frac{\lambda_2 u }{\lambda_2-2}\dir{ u} + 
	\frac{\lambda_2(\beta+1) + 2(\beta-1) }{(\beta-1)(\lambda_2-2)} \press\,\dir{ \press}  + \\
	&\frac{\lambda_2(3\beta-1) + 2(\beta-1)  }{(\beta-1)(\lambda_2-2)}\dens\,\dir{ \dens} -
	\frac{2\beta\lambda_2 \entr }{(\beta-1)(\lambda_2-2)}\,\dir{ \entr}
	+ \frac{2\lambda_2 \temp }{(\beta-1)(\lambda_2-2)}\dir{ \temp}   
	\end{aligned}$\\
	\hline
	$h(a)= \lambda_1e^{\lambda_2a}$, $\lambda_2\neq 0 $
	&$\begin{aligned} 
	&\dir{ t}, \quad
	\dir{\press }, \quad 
	\dir{\entr}, \quad \\
	&t\,\dir{ t}-\frac{2	}{\lambda_2}\,\dir{a} - u\,\dir{ u}  +\frac{\beta+1}{\beta-1} \press\,\dir{ \press}  +
	\frac{3\beta-1 }{\beta-1}\dens\,\dir{ \dens} -
	\frac{2\beta\entr}{\beta-1} \dir{ \entr} +
	\frac{2\temp  }{\beta-1}\dir{ \temp}
	\end{aligned}$\\
	\hline
	$h(a)= \ln{a}$ 
	&$\begin{aligned} 
	&\dir{ t}, \quad
	\dir{\press }, \quad 
	\dir{\entr}, \quad 
	 t\,\dir{ t}+ a  \,\dir{a}- \press\,\dir{ \press} -\dens\,\dir{\dens}
	\end{aligned}$\\
	\hline
\end{tabular}
\vspace{12pt}
\endgroup

\section{Thermodynamic states with a one-dimensional symmetry algebra} \label{sec:therm}

Recall that our approach to thermodynamics is based on geometric interpretation
of thermodynamic states as Lagrangian manifolds \cite{DLTwisla}.

In this section we study thermodynamic states that admit a one-dimensional
symmetry algebra
\[
Z=\gamma_1Y_1+ \gamma_2Y_2+ \ldots + \gamma_kY_k,
\]
where $Y_i$ are pure thermodynamic symmetries, that is, elements of the algebra
$\LieAlgebra{h}$. The cases of thermodynamic states admitting a two-dimensional
symmetry algebra can be studied in the similar manner.

Following our approach, we come to a system on the Lagrangian manifold $L$
\[
\left\{
\begin{aligned}
&\Omega\vert_{L} =0, \\
&(\iota_{Z}{\Omega})\vert_{L} = 0
\end{aligned} \right.
\]
with the condition $\kappa\vert_{L}<0$.

The latter is equivalent to an overdetermined PDE system on the specific energy
$\energy = \energy(\dens,\entr)$. We can obtain two state equations from its
solution with relations
\[
\temp = \energy_{\entr}, \quad 
\press = \dens^2\energy_{\dens}.
\]

\subsection{$\zeta(\temp)$ is an arbitrary function}

First of all, since the quadratic form $\kappa$ is degenerated everywhere, there
are no valid thermodynamic states admitting a one-dimensional symmetry algebra
for an arbitrary function $h(a)$ and $h(a)=\lambda a^2$.

Note that for $h(a)=const$, $h(a)=\lambda a$ and $h(a)=\ln a$ the pure
thermodynamic symmetries $\LieAlgebra{h}$ are same. 

If a one-dimensional symmetry subalgebra is generated by
\[
\gamma_1Y_1+\gamma_2Y_2+\gamma_3Y_3 = \gamma_1 \dir{\press} +\gamma_2\dir{\entr}
+ \gamma_3(\press\,\dir{ \press} + \dens\,\dir{ \dens}),
\]
then the corresponding thermodynamic state (or Lagrangian manifold) is defined
by the equations
\[
\press = \frac{-(\gamma_2 F^{\prime} + C )\dens}{\gamma_3} -
\frac{\gamma_1}{\gamma_3}, \quad
\temp = F^{\prime}, \quad
F=F\left( \entr- \frac{\gamma_2}{\gamma_3}\ln \dens \right) 
\]
where $C$ is a constant and $F$ is an arbitrary function.
The admissibility condition $\kappa\vert_{L}<0$ leads to the relations 
\[
F^{\prime}>0, \quad  F^{\prime\prime}>0, \quad
\frac{\gamma_2 F^{\prime} + C}{\gamma_3}<0
\]
for all $\entr \in(-\infty,\entr_0]$.

\subsection{$\zeta(\temp)= \alpha \temp$}

\textbf{1.} {$h(a)$ is arbitrary, $h(a)=\lambda a^2$}

If a one-dimensional symmetry subalgebra is generated by
\begin{equation*} 
\gamma_1Y_1 + \gamma_2Y_{2} + \gamma_3Y_{3} = \gamma_1 \dir{ \press} +\gamma_2
\dir{ \entr} +\gamma_3(  \press\,\dir{ \press} + \dens\,\dir{ \dens} -
\entr\,\dir{ \entr} + \temp\,\dir{ \temp}),
\end{equation*}
then the corresponding thermodynamic state is defined by the equations
\[
\press = \dens^2\left( \entr- \frac{\gamma_2}{\gamma_3} \right)  F^{\prime} +
C\dens - \frac{\gamma_1}{\gamma_3}, \quad
\temp = \dens F^{\prime},\quad
F=F\left( \left( \entr- \frac{\gamma_2}{\gamma_3}  \right) \dens\right) ,
\]
where $C$ is a constant and $F$ is an arbitrary function.
The admissibility condition $\kappa\vert_{L}<0$ leads to the relations 
\[
F^{\prime}>0, \quad  F^{\prime\prime}>0, \quad
C F^{\prime\prime} - (F^{\prime})^2 >0.
\]
\textbf{2.} {$h(a)=const$, $h(a)=\lambda a$, $h(a)=\ln a$}

If a one-dimensional symmetry subalgebra is generated by
\begin{equation*} 
\gamma_1Y_1 + \gamma_2Y_{2} + \gamma_3 Y_{3} +\gamma_4 Y_4 = \gamma_1\dir{ \press} +\gamma_2 \dir{ \entr} +\gamma_3( \press\,\dir{ \press} + \dens\,\dir{ \dens}) + \gamma_4( \entr\,\dir{ \entr} - \temp\,\dir{ \temp} ),
\end{equation*}
then the corresponding thermodynamic state is defined by the equations
\[
\press = C\dens - \frac{F^{\prime}(\gamma_4\entr+\gamma_2)  \dens^{-\frac{\gamma_4}{\gamma_3}+1}+\gamma_1}{\gamma_3} ,\quad
\temp = \dens^{-\frac{\gamma_4}{\gamma_3}} F^{\prime},\quad
F=F\left( \left( \entr+ \frac{\gamma_2}{\gamma_4}  \right) \dens^{-\frac{\gamma_4}{\gamma_3}}\right),
\]
where $C$ is a constant and $F$ is an arbitrary function.
The admissibility condition $\kappa\vert_{L}<0$ leads to the relations 
\[
F^{\prime}>0, \quad  F^{\prime\prime}>0, \quad
F^{\prime}F^{\prime\prime}(\gamma_3+\gamma_4)(\gamma_4\entr+\gamma_2)\dens^{-\frac{\gamma_4}{\gamma_3}} - C\gamma_3^2F^{\prime\prime} + \gamma_4^2(F^{\prime})^2<0
\]
for all $\entr \in(-\infty,\entr_0]$.

\subsection{$\zeta(\temp)= \alpha \temp^\beta$, $\beta\neq 1$}

Since the quadratic form $\kappa$ is degenerated everywhere, there are no valid
thermodynamic states admitting a one-dimensional symmetry algebra for an
arbitrary function $h(a)$.

\medskip
\textbf{1.} $h(a)= const$, {$h(a)= \lambda a$,  $\lambda\neq 0$} 

If a one-dimensional symmetry subalgebra is generated by
\begin{equation*} 
\gamma_1 Y_1 +\gamma_2Y_2 +\gamma_3Y_3 +\gamma_4Y_4=\gamma_1\,\dir{ \press} + \gamma_2\,\dir{ \entr} +  
\gamma_3(\press\,\dir{ \press}+\dens\,\dir{ \dens}) +
\gamma_4\left( (\beta-1)\dens\,\dir{ \dens} - \beta \entr\,\dir{ \entr} + \temp\,\dir{ \temp}\right)  ,
\end{equation*}
then the corresponding thermodynamic state is defined by the equations
\[
\press = \frac{F^{\prime}(\beta\gamma_4\entr-\gamma_2)  \dens^{\frac{\beta\gamma_4+\gamma_3}{\gamma_3+(\beta-1)\gamma_4}}-F\gamma_4(\beta-1)
\dens^{\frac{\gamma_3}{\gamma_3+(\beta-1)\gamma_4}}}{\gamma_3+(\beta-1)\gamma_4}-\frac{\gamma_1}{\gamma_3} ,\quad
\temp = F^{\prime}\dens^{\frac{\gamma_4}{\gamma_3+(\beta-1)\gamma_4}} ,\quad
\]
\[
F=F\left( \left( \entr- \frac{\gamma_2}{\beta\gamma_4} 
 \right)\dens^{\frac{\beta\gamma_4}{\gamma_3+(\beta-1)\gamma_4}} \right)
\]
where $F$ is an arbitrary function.
The admissibility condition $\kappa\vert_{L}<0$ leads to the relations
\[
F^{\prime}>0, \quad  F^{\prime\prime}>0, \quad
F^{\prime}F^{\prime\prime}(\gamma_3-\gamma_4)(\beta\gamma_4\entr-\gamma_2)\dens^{\frac{\beta\gamma_4}{\gamma_3+(\beta-1)\gamma_4}} - \gamma_4(\gamma_3(\beta-1)FF^{\prime\prime} + \gamma_4F^{\prime 2})>0
\]
for all $\entr \in(-\infty,\entr_0]$.

\medskip
\textbf{2.} {$h(a)= \lambda a^2$,  $\lambda\neq 0$}

If a one-dimensional symmetry subalgebra is generated by
\begin{equation*} 
\gamma_1Y_1 + \gamma_2Y_{2} +\gamma_3Y_{3} =\gamma_1 \dir{ \press} +\gamma_2\dir{ \entr} +\gamma_3 \left( \beta\press\,\dir{ \press}+ (2\beta-1)\dens\,\dir{ \dens} - \beta \entr\,\dir{ \entr} + \temp\,\dir{ \temp} \right) ,
\end{equation*}
then the corresponding thermodynamic state is defined by the equations
\[
\press = \frac{F^{\prime}(\beta\gamma_3\entr-\gamma_2)  \dens^{\frac{2\beta}{2\beta-1}}-
F\gamma_3(\beta-1)
\dens^{\frac{\beta}{2\beta-1}}}{\gamma_3(2\beta-1)}-
\frac{\gamma_1}{\beta\gamma_3} ,\quad
\temp = F^{\prime}\dens^{\frac{1}{2\beta-1}} ,\quad
F=F\left( \left( \entr- \frac{\gamma_2}{\beta\gamma_3} 
\right)\dens^{\frac{\beta}{2\beta-1}} \right),
\]
where $F$ is an arbitrary function.
The admissibility condition $\kappa\vert_{L}<0$ leads to the relations
\[
F^{\prime}>0, \quad  F^{\prime\prime}>0, \quad
F^{\prime}F^{\prime\prime}(\beta-1)(\beta\entr-\frac{\gamma_2}{\gamma_3})
\dens^{\frac{\beta}{2\beta-1}} - \beta(\beta-1)FF^{\prime\prime} - F^{\prime 2}>0
\]
for all $\entr \in(-\infty,\entr_0]$.

If $\beta=1/2$ the thermodynamic state is defined by the equations
\[
\press = \frac{\dens^2F^{\prime}(\dens) + 2\gamma_1\entr }{2\gamma_2-\gamma_3\entr} ,\quad
\temp = \frac{\gamma_3\dens F(\dens) -4\gamma_2(C\dens+\gamma_1)}{\dens(2\gamma_2-\gamma_3\entr)^2},
\]
where $C$ is a constant and $F$ is an arbitrary function.
The admissibility condition gives 
\[
\frac{2\gamma_2-\gamma_3\entr}{\gamma_3} >0, \quad
(\press^2+4\press\dens\entr\temp)\gamma_3^2 - ((2\dens^3F^{\prime\prime}+8(\gamma_2\press-\gamma_1\entr))\dens\temp-4\gamma_1\press)\gamma_3 + 4\gamma_1^2<0.
\]

\medskip
\textbf{3.} $h(a)= \lambda_1a^{\lambda_2}$, $\lambda_2\neq 0,1,2$ 

If a one-dimensional symmetry subalgebra is generated by
\begin{align*} 
&\gamma_1Y_1 + \gamma_2Y_{2} +\gamma_3Y_{3} =\gamma_1 \dir{ \press} +\gamma_2\dir{ \entr} + \\
&\quad \gamma_3 \left( \left( \lambda_2(\beta+1) + 2(\beta-1)\right) \press\,\dir{ \press}+ (\lambda_2(3\beta-1) + 2(\beta-1))\dens\,\dir{ \dens} - 2\beta\lambda_2 \entr\,\dir{ \entr} +2\lambda_2 \temp\,\dir{ \temp}\right)  ,
\end{align*}
then the corresponding thermodynamic state is defined by the equations
\[
\press = \frac{(2\beta\lambda_2\gamma_3\entr -\gamma_2) F^{\prime}\dens^{\frac{2\lambda_2}{A}+1} -
2(\beta-1)\lambda_2\gamma_3F\dens^{\frac{2\lambda_2(1-\beta)}{A}+1} }{A\gamma_3} - \frac{\gamma_1}{\gamma_3(\lambda_2(\beta+1) +2(\beta-1))}, \quad
\]
\[
\temp = \dens^{\frac{2\lambda_2}{A}}F^{\prime},\quad
F = F\left( \dens^{\frac{2\beta\lambda_2}{A}}\left( \entr - \frac{\gamma_2}{2\beta\lambda_2\gamma_3}\right)    \right),
\]
where $A={\lambda_2(3\beta-1) + 2(\beta-1)}$ 
and the admissibility condition $\kappa\vert_{L}<0$ gives 
\[
F^{\prime}>0, \quad  F^{\prime\prime}>0, \quad
\]
\[
\left( 2\beta\lambda_2\entr - \frac{\gamma_2}{\gamma_3}\right) \left(\beta-1 \right)(\lambda_2+2) F^{\prime}F^{\prime\prime}\dens^{\frac{2\beta\lambda_2}{A}}
-2\lambda_2\left( (\lambda_2(\beta+1)+2(\beta-1))(\beta-1)FF^{\prime\prime}+2\lambda_2F^{\prime 2} \right)  >0.
\]

\medskip
\textbf{4.} $h(a)=\lambda_1e^{\lambda_2a}, \, \lambda_2\neq 0$

If a one-dimensional symmetry subalgebra is generated by
\[
\gamma_1Y_1 + \gamma_2Y_{2} +\gamma_3Y_{3} =\gamma_1 \dir{ \press} +\gamma_2\dir{ \entr} +\gamma_3 \left( (\beta+1) \press\,\dir{ \press}+ (3\beta-1) \dens\,\dir{ \dens} - 2\beta \entr\,\dir{ \entr} +2 \temp\,\dir{ \temp} \right),
\]
then the corresponding thermodynamic state is defined by the equations
\[
\press = \frac{(2\beta\gamma_3\entr-\gamma_2)\dens^{\frac{3\beta+1}{3\beta-1}}F^{\prime}-
2\gamma_3(\beta-1)\dens^{\frac{\beta+1}{3\beta-1}}F}{\gamma_3(3\beta-1)} - \frac{\gamma_1}{\gamma_3(\beta+1)}, \quad
\temp = \dens^{\frac{2}{3\beta-1}}F^{\prime}
\]
\[
F = F\left( \dens^{\frac{2\beta}{3\beta -1}}\left( \entr - \frac{\gamma_2}{2\beta\gamma_3}\right)    \right)
\]
and
\[
F^{\prime}>0, \quad  F^{\prime\prime}>0, \quad
(\beta-1)\left(2\beta\entr - \frac{\gamma_2}{\gamma_3} \right) F^{\prime}F^{\prime\prime}\dens^{\frac{2\beta}{3\beta-1}} -
2(\beta^2-1)FF^{\prime\prime} -4 F^{\prime 2}>0 .
\]

If $\beta=1/3$ then  the thermodynamic state is defined by the equations 
\[
\press = \frac{\dens^2F^{\prime}(\dens) - 3\gamma_1\entr (\gamma_3\entr - 3\gamma_2)}{(2\gamma_3\entr - 3\gamma_2)^2}, \quad
\temp = \frac{ 27\gamma_2^2(C\dens+\gamma_1) - 4\gamma_3\dens F(\dens) }{\dens (2\gamma_3\entr - 3\gamma_2)^3},
\]
and the admissibility condition $\kappa\vert_{L}<0$ gives
\[
\frac{6\gamma_3}{2\gamma_3\entr - 3\gamma_2}<0, \qquad
\frac{6\gamma_3(2F^{\prime}+\dens F^{\prime\prime})}{2\gamma_3\entr - 3\gamma_2} + \frac{\left(4\gamma_3\dens^2F^{\prime}+27\gamma_1\gamma_2^2 \right) ^2}{\dens^3\temp \left( 2\gamma_3\entr - 3\gamma_2\right)^4 }<0
\]
for all $\entr \in(-\infty,\entr_0]$.

If $\beta=-1$ then the thermodynamic state is defined by the equations
\[
\press = { \dens^{\frac{1}{2}} F^{\prime}}  \left( \frac{\entr}{2} + \frac{\gamma_2}{4\gamma_3} \right) - \frac{\gamma_1}{4\gamma_3} (\ln \dens - 1)  -F , \quad
\temp =  \dens^{-\frac{1}{2}} F^{\prime} , \quad
F= F\left( \dens^{\frac{1}{2}} \left( \entr + \frac{\gamma_2}{2\gamma_3} \right) \right) ,
\]
where 
\[
F^{\prime}>0, \quad  F^{\prime\prime}>0, \quad 
 \left( \entr + \frac{\gamma_2}{2\gamma_3}\right) \dens^{\frac{1}{2}} F^{\prime} F^{\prime\prime} -  F^{\prime 2} - \frac{\gamma_1}{\gamma_3}  F^{\prime\prime} >0 
\]
for all $\entr \in(-\infty,\entr_0]$.

\medskip
\textbf{5.} $h(a)= \ln a$

In this case the pure thermodynamic part of the symmetry algebra coincides with the thermodynamic part of the case when $\zeta(\temp)$ is an arbitrary function.


\section{ Differential invariants} \label{sec:invs}

In this section we recollect the notions of
kinematic and Navier--Stokes differential invariants.

As in \cite{DLTwisla}, we consider
the prolonged group actions generated
by the Lie algebras $\LieAlgebra{g_{m}}$ and 
$\LieAlgebra{g_{sym}}$ on the Navier--Stokes system $\systemEk{}$.


We call \cite{DLTwisla} a function $J$ on the manifold $\systemEk{k}$
a \textit{kinematic differential invariant of order} $\leq k$ if 
\begin{enumerate}
	\item $J$ is a rational function along fibers of the projection $\pi_{k,0}:\systemEk{k}\rightarrow \systemEk{0}$,
	\item $J$ is invariant with respect to the prolonged action of the Lie algebra $\LieAlgebra{g_{m}}$, i.e., for all $X\in \LieAlgebra{g_{m}}$,
	\begin{equation} \label{dfinv}
	X^{(k)}(J)=0,
	\end{equation}
\end{enumerate}
where $\systemEk{k}$ is the prolongation of the system $\systemEk{}$ to $k$-jets,
and $X^{(k)}$ is the $k$-th prolongation of a vector field $X\in \LieAlgebra{g_{m}}$.

Note that fibers of the projection $\systemEk{k}\rightarrow \systemEk{0}$ are irreducible algebraic manifolds.

A kinematic invariant is \textit{a Navier--Stokes invariant} if condition \eqref{dfinv} holds for all $X \in \LieAlgebra{g_{sym}}$.

We say that a point $x_k\in \systemEk{k}$ and its $\LieAlgebra{g_{m}}$-orbit $\mathcal{O}(x_k)$ (or $\LieAlgebra{g_{sym}}$-orbit) are \textit{regular}, if there are exactly $\mathrm{codim}\, \mathcal{O}(x_k) $ independent kinematic  (or Navier--Stokes) invariants in a neighborhood of this orbit. 
Otherwise, the point and the corresponding orbit are \textit{singular}.

Since the Navier--Stokes system and 
the symmetry algebras $\LieAlgebra{g_{m}}$ and
$\LieAlgebra{g_{sym}}$ satisfy 
the Lie--Tresse theorem (see \cite{KL}), and, therefore,
the kinematic and Navier--Stokes differential invariants separate
regular $\LieAlgebra{g_{m}}$- and
$\LieAlgebra{g_{sym}}$-orbits on the Navier--Stokes system $\systemEk{}$ correspondingly. 

We call a total derivative
\[
A\totalDiff{t}+B\totalDiff{a}
\]
$\LieAlgebra{g_{m}}$- or $\LieAlgebra{g_{sym}}$-invariant, if it
commutes with the prolonged action of algebra $\LieAlgebra{g_{m}}$  or $\LieAlgebra{g_{sym}}$, and $A$, $B$ are rational functions
along fibers of the projection $\pi_{k,0}:\systemEk{k}\rightarrow \systemEk{0}$ for some $k\geq 0$.

\subsection{Kinematic invariants}
\begin{theorem} 
	\begin{enumerate}
		\item The kinematic invariants field is generated
		by first-order basis differential invariants and by basis invariant derivatives. This field separates regular orbits.
		\item For the cases of arbitrary $h(a)$, as well as for  $h(a)=\lambda_1a^{\lambda_2}$, $h(a)=\lambda_1e^{\lambda_2a}$ and $h(a)=\ln a$, the basis differential invariants are
		\[
		a,\quad u,\quad \dens,\quad\entr,\quad  u_t,\quad u_a,\quad\dens_a,\quad \entr_t, \quad\entr_a,
		\]
		and the basis invariant derivatives are
		\[
		\totalDiff{t} ,\quad \totalDiff{a} . 
		\]
		\item For the cases $h(a)=const$, $h(a)=\lambda a$ 
		the basis differential invariants are
		\[
		\dens,\quad\entr,\quad u_a, \quad u_t+u u_a, \quad\dens_a, \quad \entr_a, \quad \entr_t+u \entr_a,
		\]
		and basis invariant derivatives are
		\[
		\totalDiff{t} + u \totalDiff{a} ,\quad  \totalDiff{a}.
		\]	
		\item For the case $h(a)=\lambda a^2$  
		the basis differential invariants are
		\[
		\dens,\quad\entr,\quad u_a, \quad u_t+u u_a-2\lambda g a, 
		\quad\dens_a, \quad \entr_a, \quad \entr_t+u \entr_a ,
		\]
		and basis invariant derivatives are
		\[
		\totalDiff{t} + u \totalDiff{a} ,\quad  \totalDiff{a}.
		\]	
		\item The number of independent invariants of pure order $k$	is equal to $5$ for $k\geq 1$.
	\end{enumerate}
\end{theorem}

\subsection{Navier--Stokes invariants}

Let the thermodynamic state admit a one-dimensional symmetry algebra generated by the vector field $A$.

We get basis first-order Navier--Stokes differential invariants
finding first integrals of an action of the vector $A$ on the field of kinematic invariants.

Below we list basis invariants depending on the form of functions $\zeta(\temp)$ and $h(a)$.

\subsubsection{$\zeta(\temp)$ is an arbitrary function}

\medskip
\textbf{1.} $h(a)=const$

If the thermodynamic state admits a one-dimensional symmetry algebra generated by 
\[
\xi_1 X_2 + \xi_2 X_3 + \xi_3 X_6  =
\xi_1\dir{ \press}  + \xi_2 \dir{ \entr} + \xi_3 (t\,\dir{ t}+a\,\dir{a}-\press\,\dir{ \press} - \dens\,\dir{ \dens}),
\]
then the field of Navier--Stokes differential invariants is generated by the first-order differential invariants
\[
\entr + \frac{\xi_2}{\xi_3} \ln \dens, \quad
\frac{u_a}{\dens}, \quad
\frac{u_t+uu_a}{\dens},\quad
\frac{\dens_a}{\dens^2}, \quad
\frac{\entr_a}{\dens}, \quad
\frac{\entr_t+u\entr_a}{\dens}
\]
and by the invariant derivatives 
\[
\dens^{-1}\left( \totalDiff{t} +u \totalDiff{a}\right)   ,\quad 
\dens^{-1} \totalDiff{a} .
\]

\medskip
\textbf{2.}  $h(a)=\lambda a$, $\lambda\neq 0$

If the thermodynamic state admits a one-dimensional symmetry algebra generated by
\[
\xi_1 X_2 + \xi_2 X_3 + \xi_3 X_6  =
\xi_1\dir{ \press}  + \xi_2 \dir{ \entr} + \xi_3 \left(t\,\dir{ t}+ \left( \frac{\lambda g t^2}{2}+a\right) \,\dir{ a} +\lambda g t\,\dir{ u} -\press\,\dir{ \press} - \dens\,\dir{ \dens} \right) ,
\]
then the field of Navier--Stokes differential invariants is generated by the first-order differential invariants
\[
\entr + \frac{\xi_2}{\xi_3} \ln \dens, \quad
\frac{u_a}{\dens}, \quad
\frac{u_t+uu_a-\lambda g}{\dens},\quad
\frac{\dens_a}{\dens^2}, \quad
\frac{\entr_a}{\dens}, \quad
\frac{\entr_t+u\entr_a}{\dens}
\]
and by the invariant derivatives 
\[
\dens^{-1}\left( \totalDiff{t} +u   \totalDiff{a}\right) ,\quad 
\dens^{-1} \totalDiff{a} .
\]

\medskip
\textbf{4.}  $h(a)=\ln a$

If the thermodynamic state admits a one-dimensional symmetry algebra generated by
\[
\xi_1 X_2 + \xi_2 X_3 + \xi_3 X_4  =
\xi_1\dir{ \press}  + \xi_2 \dir{ \entr} + \xi_3 \left(t\,\dir{ t}+ a \,\dir{ a} -\press\,\dir{ \press} - \dens\,\dir{ \dens} \right) ,
\]
then the field of Navier--Stokes differential invariants is generated by the first-order differential invariants
\[
\entr - \frac{\xi_2}{\xi_3} \ln a, \quad
u, \quad
a\dens,\quad
a u_t, \quad
au_a, \quad
a^2 \dens_a, \quad
a\entr_t, \quad a\entr_a
\]
and by the invariant derivatives 
\[
\dens^{-1}\totalDiff{t}   ,\quad 
\dens^{-1} \totalDiff{a} .
\]


\subsubsection{$\zeta(\temp)= \alpha \temp$}

First of all, if $h(a)$ is an arbitrary function and if the thermodynamic state admits a one-dimensional symmetry algebra generated by
\[
\xi_1 X_2 + \xi_2 X_3 + \xi_3 X_4  =
\xi_1\dir{ \press}  + \xi_2 \dir{ \entr} + \xi_3 (\press\,\dir{ \press} + \dens\,\dir{ \dens} - \entr\,\dir{ \entr} + \temp\,\dir{ \temp}),
\]
then the field of Navier--Stokes differential invariants is generated by the first-order differential invariants
\[
a, \quad
u, \quad
\left( \entr - \frac{\xi_2}{\xi_3}\right) \dens, \quad
 u_t, \quad
u_a, \quad
\frac{\dens_a}{\dens}, \quad
\dens\entr_t, \quad \dens\entr_a
\]
and by the invariant derivatives 
\[
\totalDiff{t}   ,\quad 
\totalDiff{a} .
\]

\medskip
\textbf{1.} $h(a)=const$

If the thermodynamic state admits a one-dimensional symmetry algebra generated by
\[
\xi_1 X_2 + \xi_2 X_3 + \xi_3 X_4 + \xi_4 X_7  =
\xi_1\dir{ \press}  + \xi_2 \dir{ \entr} + \xi_3 (\press\,\dir{ \press} + \dens\,\dir{ \dens} - \entr\,\dir{ \entr} + \temp\,\dir{ \temp}) + \xi_4 (t\,\dir{ t}+a\,\dir{a}-\press\,\dir{ \press} - \dens\,\dir{ \dens}),
\]
then the field of Navier--Stokes differential invariants is generated by the first-order differential invariants
\[
\frac{\dens}{u_a} \left( \entr - \frac{\xi_2}{\xi_3} \right)  , \quad
{u_a}\dens^{\frac{\xi_4}{\xi_3-\xi_4}}, \quad
\frac{u_t+uu_a}{u_a},\quad
\frac{\dens_a}{\dens u_a}, \quad
\frac{\dens\entr_a}{u_a^2}, \quad
\frac{\dens(\entr_t+u\entr_a)}{u_a^2}
\]
and by the invariant derivatives 
\[
\dens^{\frac{\xi_4}{\xi_3-\xi_4}}\left( \totalDiff{t} +u \totalDiff{a}\right)   ,\quad 
\dens^{\frac{\xi_4}{\xi_3-\xi_4}} \totalDiff{a} .
\]

\medskip
\textbf{2.}  $h(a)=\lambda a$, $\lambda\neq 0$

If the thermodynamic state admits a one-dimensional symmetry algebra generated by
\begin{align*}
&\xi_1 X_2 + \xi_2 X_3 + \xi_3 X_4 + \xi_4 X_7  =
\xi_1\dir{ \press}  + \xi_2 \dir{ \entr} +\\
&\quad \qquad \xi_3 (\press\,\dir{ \press} + \dens\,\dir{ \dens} - \entr\,\dir{ \entr} + \temp\,\dir{ \temp}) + \xi_4 \left(t\,\dir{ t}+ \left( \frac{\lambda g t^2}{2}+a\right) \,\dir{ a} +\lambda g t\,\dir{ u} -\press\,\dir{ \press} - \dens\,\dir{ \dens} \right) ,
\end{align*}
then the field of Navier--Stokes differential invariants is generated by the first-order differential invariants
\[
\frac{\dens}{u_a}\left( \entr - \frac{\xi_2}{\xi_3}\right) , \quad
{u_a}\dens^{\frac{\xi_4}{\xi_3-\xi_4}}, \quad
\frac{u_t+uu_a-\lambda g}{\dens},\quad
\frac{\dens_a}{\dens u_a}, \quad
\frac{\dens\entr_a}{u_a^2}, \quad
\frac{\dens(\entr_t+u\entr_a)}{u_a^2}
\]
and by the invariant derivatives 
\[
\dens^{\frac{\xi_4}{\xi_3-\xi_4}}\left( \totalDiff{t} +u   \totalDiff{a}\right) ,\quad 
\dens^{\frac{\xi_4}{\xi_3-\xi_4}} \totalDiff{a} .
\]

\medskip
\textbf{3.}  $h(a)=\lambda a^2$, $\lambda\neq 0$

If the thermodynamic state admits a one-dimensional symmetry algebra generated by
\begin{align*}
&\xi_1 X_2 + \xi_2 X_3 + \xi_3 X_4   =
\xi_1\dir{ \press}  + \xi_2 \dir{ \entr} + \xi_3 (\press\,\dir{ \press} + \dens\,\dir{ \dens} - \entr\,\dir{ \entr} + \temp\,\dir{ \temp}) ,
\end{align*}
then the field of Navier--Stokes differential invariants is generated by the first-order differential invariants
\[
\left( \entr - \frac{\xi_2}{\xi_3}\right) \dens, \quad
u_t + uu_a - 2\lambda ga, \quad
u_a, \quad
\frac{\dens_a}{\dens}, \quad
\dens\left( \entr_t +u\entr_a\right) , \quad \dens\entr_a
\]
and by the invariant derivatives 
\[
\totalDiff{t} +u \totalDiff{a} ,\quad 
\totalDiff{a} .
\]

\medskip
\textbf{4.}  $h(a)=\ln a$

If the thermodynamic state admits a one-dimensional symmetry algebra generated by
\[
\xi_1 X_2 + \xi_2 X_3 + \xi_3 X_4 + \xi_4 X_5  =
\xi_1\dir{ \press}  + \xi_2 \dir{ \entr} + \xi_3 (\press\,\dir{ \press} + \dens\,\dir{ \dens} - \entr\,\dir{ \entr} + \temp\,\dir{ \temp}) + \xi_4 \left(t\,\dir{ t}+ a \,\dir{ a} -\press\,\dir{ \press} - \dens\,\dir{ \dens} \right) ,
\]
then the field of Navier--Stokes differential invariants is generated by the first-order differential invariants
\[
u, \quad
\dens a^{1-\frac{\xi_3}{\xi_4}},\quad
a\dens \left( \entr - \frac{\xi_2}{\xi_3}\right)  , \quad
a u_t, \quad
au_a, \quad
\frac{a \dens_a}{\dens } , \quad
a^2\dens\entr_t, \quad a^2\dens\entr_a
\]
and by the invariant derivatives 
\[
\dens^{\frac{\xi_4}{\xi_3-\xi_4}}\totalDiff{t}   ,\quad 
\dens^{\frac{\xi_4}{\xi_3-\xi_4}} \totalDiff{a} .
\]

\subsubsection{$\zeta(\temp)= \alpha \temp^\beta$, $\beta\neq 1$}

\medskip
\textbf{1.} $h(a)=const$

If the thermodynamic state admits a one-dimensional symmetry algebra generated by
\begin{align*}
&\xi_1 X_2 + \xi_2 X_3 + \xi_3 X_6 + \xi_4 X_7 = \xi_1\dir{ \press}  + \xi_2 \dir{ \entr} + \xi_3  \left(t\,\dir{ t}+ a \,\dir{ a} -\press\,\dir{ \press} - \dens\,\dir{ \dens} \right) +\\
&\quad\qquad \qquad \qquad \xi_4\left( a\,\dir{ a} + u\,\dir{ u} 
-\frac{2\beta}{\beta-1} \press\,\dir{ \press}  -
\frac{4\beta-2  }{\beta-1}\dens\,\dir{ \dens} +
\frac{2\beta}{\beta-1} \entr\,\dir{ \entr} -
\frac{2  }{\beta-1}\temp\dir{ \temp}  \right) ,
\end{align*}
then the field of Navier--Stokes differential invariants is generated by the first-order differential invariants
\[
\frac{\dens^3u_a}{\dens_a^2} \left( \entr + \frac{\xi_2(\beta-1)}{2\xi_4\beta} \right)  , \quad
{u_a}\dens^{\frac{-\xi_3(\beta-1)}{\xi_3(\beta-1)+2\xi_4(2\beta -1)}}, \quad
(u_t+uu_a)\dens^{\frac{-(\xi_3+\xi_4)(\beta-1)}{\xi_3(\beta-1)+2\xi_4(2\beta -1)}},\quad
\frac{\dens_a(u_t+uu_a)}{\dens u_a^2}, \quad
\]
\[
\frac{\dens^4u_a\entr_a}{\dens_a^3}, \quad
\frac{\dens^3(\entr_t+u\entr_a)}{\dens_a^2}
\]
and by the invariant derivatives 
\[
\dens^{\frac{-\xi_3(\beta-1)}{\xi_3(\beta-1)+2\xi_4(2\beta -1)}}\left( \totalDiff{t} +u \totalDiff{a}\right)   ,\quad 
\dens^{\frac{-(\xi_3+\xi_4)(\beta-1)}{\xi_3(\beta-1)+2\xi_4(2\beta -1)}} \totalDiff{a} .
\]

\medskip
\textbf{2.}  $h(a)=\lambda a$, $\lambda\neq 0$

If the thermodynamic state admits a one-dimensional symmetry algebra generated by
\begin{align*}
&\xi_1 X_2 + \xi_2 X_3 + \xi_3 X_6 + \xi_4 X_7 = \xi_1\dir{ \press}  + \xi_2 \dir{ \entr} +
 \xi_4\left( t\,\dir{ t}+ \left( \frac{\lambda g t^2}{2}+a\right) \,\dir{ a} + \lambda g t\,\dir{ u}-\press\,\dir{ \press} - \dens\,\dir{ \dens} \right)  + \\
& \xi_3  \left(t\,\dir{ t}+2a\,\dir{a} +  u\,\dir{ u}  -\frac{3\beta-1}{\beta-1} \press\,\dir{ \press}  -
\frac{5\beta-3 }{\beta-1}\dens\,\dir{ \dens} +
\frac{2\beta}{\beta-1} \entr\,\dir{ \entr} -
\frac{2  }{\beta-1}\temp\dir{ \temp} \right) ,
\end{align*}
then the field of Navier--Stokes differential invariants is generated by the first-order differential invariants
\[
\frac{\dens^3u_a}{\dens_a^2} \left( \entr + \frac{\xi_2(\beta-1)}{2\xi_3\beta} \right)  , \quad
{u_a}\dens^{\frac{-(\xi_3+\xi_4)(\beta-1)}{\xi_3(5\beta-3)+\xi_4(\beta -1)}}, \quad
(u_t+uu_a-\lambda g )\dens^{\frac{-\xi_4(\beta-1)}{\xi_3(5\beta-3)+\xi_4(\beta -1)}},\quad
\]
\[
\frac{\dens_a(u_t+uu_a-\lambda g)}{\dens u_a^2}, \quad
\frac{\dens^4u_a\entr_a}{\dens_a^3}, \quad
\frac{\dens^3(\entr_t+u\entr_a)}{\dens_a^2}
\]
and by the invariant derivatives 
\[
\dens^{\frac{-(\xi_3+\xi_4)(\beta-1)}{\xi_3(5\beta-3)+\xi_4(\beta -1)}}\left( \totalDiff{t} +u \totalDiff{a}\right)   ,\quad 
\dens^{\frac{-(2\xi_3+\xi_4)(\beta-1)}{\xi_3(5\beta-3)+\xi_4(\beta -1)}} \totalDiff{a} .
\]

\medskip
\textbf{3.}  $h(a)=\lambda a^2$, $\lambda\neq 0$

If the thermodynamic state admits a one-dimensional symmetry algebra generated by
\begin{align*}
&\xi_1 X_2 + \xi_2 X_3 + \xi_3 X_6  = \xi_1\dir{ \press}  + \xi_2 \dir{ \entr} + \\
&\qquad \qquad \qquad
\xi_3  \left(a\, \dir{ a} +u\,\dir{ u}  
-\frac{2\beta}{\beta-1} \press\,\dir{ \press}  -
\frac{4\beta-2  }{\beta-1}\dens\,\dir{ \dens} +
\frac{2\beta}{\beta-1} \entr\,\dir{ \entr} -
\frac{2  }{\beta-1}\temp\dir{ \temp}  \right) ,
\end{align*}
then the field of Navier--Stokes differential invariants is generated by the first-order differential invariants
\[
\frac{\dens^3u_a}{\dens_a^2} \left( \entr + \frac{\xi_2(\beta-1)}{2\xi_3\beta} \right)  , \quad
{u_a}, \quad
(u_t+uu_a-2\lambda g a )\dens^{\frac{\beta-1}{2(2\beta -1)}},\quad
\frac{\dens_a(u_t+uu_a-2\lambda g a)}{\dens u_a^2}, \quad
\]
\[
\frac{\dens^4u_a\entr_a}{\dens_a^3}, \quad
\frac{\dens^3(\entr_t+u\entr_a)}{\dens_a^2}
\]
and by the invariant derivatives 
\[
 \totalDiff{t} +u \totalDiff{a}  ,\quad 
\dens^{-\frac{\beta-1}{2(2\beta -1)}} \totalDiff{a} .
\]

If $\beta= 1/2$ then the field of Navier--Stokes differential invariants is generated by the first-order differential invariants
\[
\dens, \quad
\frac{ 2\xi_3\entr -\xi_2}{\dens_a^2}   , \quad
{u_a}, \quad
{\dens_a(u_t+uu_a-2\lambda g a)}, \quad
\frac{\entr_a}{\dens_a^3}, \quad
\frac{\entr_t+u\entr_a}{\dens_a^2}
\]
and by the invariant derivatives 
\[
\totalDiff{t} +u \totalDiff{a}  ,\quad 
\frac{1}{\sqrt{\xi_2-2\xi_3\entr}} \totalDiff{a} .
\]

\medskip
\textbf{4.}  $h(a)= \lambda_1a^{\lambda_2}$, $\lambda\neq 0, 1, 2$

If the thermodynamic state admits a one-dimensional symmetry algebra generated by
\[
\begin{aligned}
& \xi_1 X_2 + \xi_2 X_3 + \xi_3 X_4  = \xi_1\dir{ \press}  + \xi_2 \dir{ \entr} +
\xi_3 \left( t\,\dir{ t} -\frac{2a}{\lambda_2-2} \dir{ a} - \frac{\lambda_2 u }{\lambda_2-2}\dir{ u}  -
\frac{2\beta\lambda_2  }{(\beta-1)(\lambda_2-2)}\entr\,\dir{ \entr} \right.  + \\
&\left. 
\frac{\lambda_2(\beta+1) + 2(\beta-1) }{(\beta-1)(\lambda_2-2)} \press\,\dir{ \press}+
\frac{\lambda_2(3\beta-1) + 2(\beta-1)  }{(\beta-1)(\lambda_2-2)}\dens\,\dir{ \dens}
+ \frac{2\lambda_2  }{(\beta-1)(\lambda_2-2)}\temp\,\dir{ \temp}\right)  ,
\end{aligned}
\]
then the field of Navier--Stokes differential invariants is generated by the first-order differential invariants
\[
u^2a^{-\lambda_2}, \quad
\dens u a^{1+\frac{\lambda_2\beta}{\beta-1}},\quad
au \dens  \left( \entr - \frac{\xi_2(\beta-1)(\lambda_2-2)}{2\lambda_2\beta\xi_3}\right)  , \quad
\frac{a u_t}{u^2}, \quad
\frac{au_a}{u}, \quad
\frac{a \dens_a}{\dens } , \quad
a^2\dens\entr_t, \quad a^2u\dens\entr_a
\]
and by the invariant derivatives 
\[
\dens^{\frac{(\lambda_2-2)(\beta-1)}{\lambda_2(3\beta-1) + 2(\beta-1) }}\totalDiff{t}   ,\quad 
\dens^{\frac{-2(\beta-1)}{\lambda_2(3\beta-1) + 2(\beta-1) }} \totalDiff{a} .
\]

\medskip
\textbf{5.}  $h(a)= \lambda_1e^{\lambda_2a}$

If the thermodynamic state admits a one-dimensional symmetry algebra generated by
\[
\begin{aligned}
& \xi_1 X_2 + \xi_2 X_3 + \xi_3 X_4  = \xi_1\dir{ \press}  + \xi_2 \dir{ \entr} +\\
&\qquad\qquad\qquad
\xi_3 \left(  t\,\dir{ t}-\frac{2	}{\lambda_2}\,\dir{a} - u\,\dir{ u}  +\frac{\beta+1}{\beta-1} \press\,\dir{ \press}  +  
\frac{3\beta-1 }{\beta-1}\dens\,\dir{ \dens} -
\frac{2\beta}{\beta-1} \entr\,\dir{ \entr} +
\frac{2  }{\beta-1}\temp\dir{ \temp}\right)  ,
\end{aligned}
\]
then the field of Navier--Stokes differential invariants is generated by the first-order differential invariants
\[
u^2e^{-\lambda_2a}, \quad
 u\dens  e^{\frac{\lambda_2\beta a}{\beta-1}},\quad
au \dens  \left( \entr - \frac{\xi_2(\beta-1)}{2\beta\xi_3}\right)  , \quad
\frac{u_t}{u^2}, \quad
\frac{u_a}{u}, \quad
\frac{\dens_a}{\dens } , \quad
\dens\entr_t, \quad u\dens\entr_a
\]
and by the invariant derivatives 
\[
\dens^{\frac{\beta-1}{3\beta-1 }}\totalDiff{t}   ,\quad 
 \totalDiff{a} .
\]

If $\beta=1/3$ then the field of Navier--Stokes differential invariants is generated by the first-order differential invariants
\[
u^2e^{-\lambda_2a}, \quad
\dens  ,\quad
u   \left( \entr + \frac{\xi_2}{\xi_3}\right)  , \quad
\frac{u_t}{u^2}, \quad
\frac{u_a}{u}, \quad
{\dens_a} , \quad
\entr_t, \quad u\entr_a
\]
and by the invariant derivatives 
\[
(\xi_3\entr+\xi_2)\totalDiff{t}   ,\quad 
\totalDiff{a} .
\]

\medskip
\textbf{6.}  $h(a)=\ln a$

If the thermodynamic state admits a one-dimensional symmetry algebra generated by
\[
\xi_1 X_2 + \xi_2 X_3 + \xi_3 X_4  =
\xi_1\dir{ \press}  + \xi_2 \dir{ \entr} + \xi_3 \left(t\,\dir{ t}+ a \,\dir{ a} -\press\,\dir{ \press} - \dens\,\dir{ \dens} \right) ,
\]
then the field of Navier--Stokes differential invariants is generated by the first-order differential invariants
\[
\entr - \frac{\xi_2}{\xi_3} \ln a, \quad
u, \quad
a\dens,\quad
a u_t, \quad
au_a, \quad
a^2 \dens_a, \quad
a\entr_t, \quad a\entr_a
\]
and by the invariant derivatives 
\[
\dens^{-1}\totalDiff{t}   ,\quad 
\dens^{-1} \totalDiff{a} .
\]


\section*{Appendix} \label{sec:app}

Let us define a space curve as a pair of plane curve $(x(\tau), y(\tau))$ and
function $z(\tau)$ that serves as a way of lifting the plain curve. We denote
length of the plane curve $\int\limits_0^{\tau
}\sqrt{x^2_{\theta}+y^2_{\theta}}\,d\theta$ by $l(\tau)$.

Depending on the particular form of the function $h(a)$,
we get different forms of the `lifting' function. Below,
we enumerate all six particular cases of the function $h$
arising in classification of Lie algebras
(see Section \ref{sec:symmetries}). In each case, we find
a relation between the `lifting' function $z$ and the length
of the plane curve $l(\tau)$.

\medskip
\textbf{1.} $h(a)=const$

The first way of lifting a plane curve is to translate the whole curve along $z$-axis, i.e., if $h(a)= const$ then $z(\tau)= const$.

\medskip
\textbf{2.} $h(a)= \lambda a$, $\lambda \neq 0$

The second way to lift a plane curve is lifting 
proportional to its length, i.e., if
$h(a)= \lambda a$  then the relation between the `lifting' function $z(\tau)$ and the length $ l(\tau) $ of plane projection of curve has the form
\[
z(\tau)= \pm \frac{   \lambda  }{\sqrt{1-\lambda^2}}\, l(\tau) + C ,
\] 
where $ \lambda^2<1$ and $C$ is a constant. Here, if $\lambda=\pm 1$ then $x(t)=y(t)=const$, and we have a vertical line.

\medskip
\textbf{3.} $h(a)= \lambda a^2$, $\lambda\neq 0$ 

In this case the relation between the `lifting'
function $z(\tau)$ and the length of the plane curve
is the following
\[
\sqrt{4\lambda z (1-4\lambda z)} - \arccos(\sqrt{4\lambda z}) =\pm 4\lambda l(\tau).
\]

\begin{figure}[h]
	\centering
	\begin{subfigure}[b]{0.36\textwidth}
		\includegraphics[width=\textwidth]{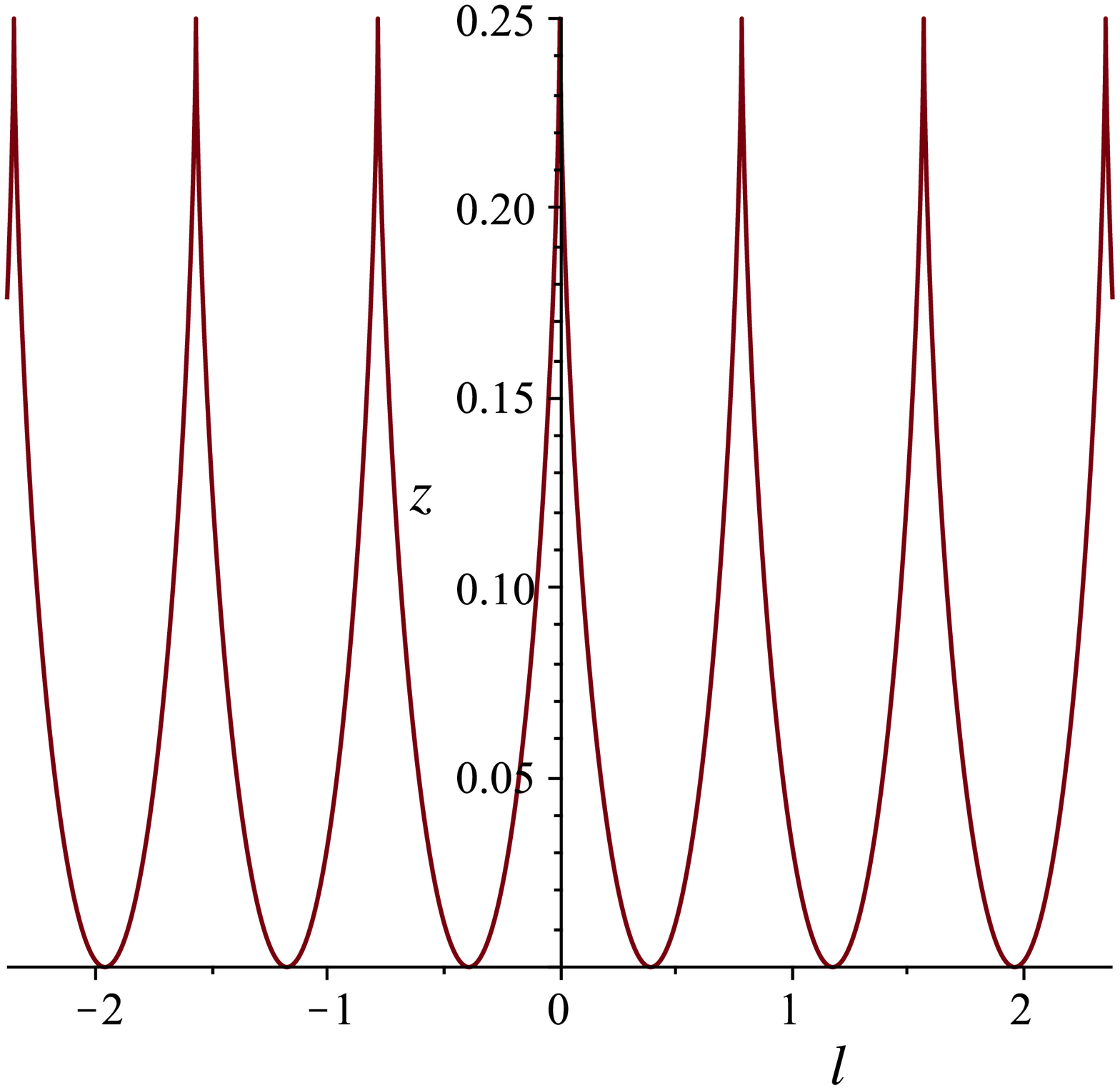}
		\caption{}\label{fig:pic3-1}
	\end{subfigure}  \qquad
	\begin{subfigure}[b]{0.40\textwidth}
		\centering
		\includegraphics[width=\textwidth]{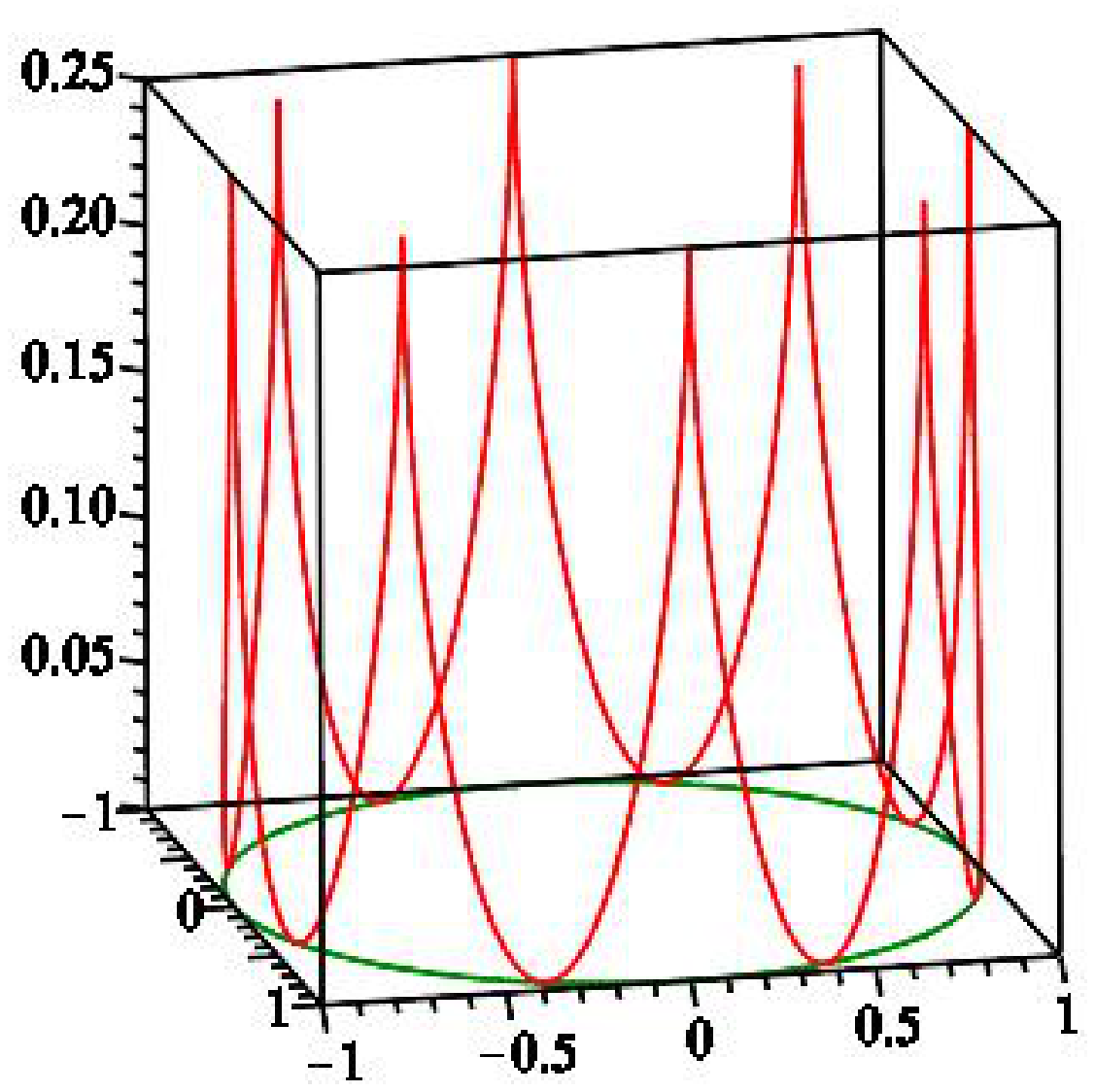}
		\caption{}\label{fig:pic3-2}
	\end{subfigure}
	\caption{}
\end{figure}

This relation is presented in Figure \ref{fig:pic3-1}.
An example of lifting of a unit circle with
the function is given in Figure \ref{fig:pic3-2}.

\medskip
\textbf{4.} $h(a)= \lambda_1a^{\lambda_2} $, $\lambda_2\neq 0,1,2$

In this case, relation between the `lifting' function $z(\tau)$
and the length of the plane curve is rather complex
and involves hypergeometric functions. For example, for the case $\lambda_1=1$, $\lambda_2=\frac{11}{3}$ we get

\[
z^{\frac{3}{11}}\, _2F_1\left(-\frac{1}{2},\frac{3}{16};\frac{19}{16};
\frac{121}{9}z^{\frac{16}{11}}\right) = \pm l(\tau).
\]

\medskip
\textbf{5.} $h(a)= \lambda_1e^{\lambda_2a} $  

The relation between the `lifting' function $z(\tau)$
and the length of the plane curve is
\[
\sqrt{1-\lambda_2^2z^2}-\frac{1}{2}\ln\frac{1+\sqrt{1-\lambda_2^2z^2}}{1-\sqrt{1-\lambda_2^2z^2}} =\pm \lambda_2 l(\tau),
\]
where $\lambda_2^2z^2<1$.

\begin{figure}[h]
	\centering
	\begin{subfigure}[b]{0.38\textwidth}
		\includegraphics[width=\textwidth]{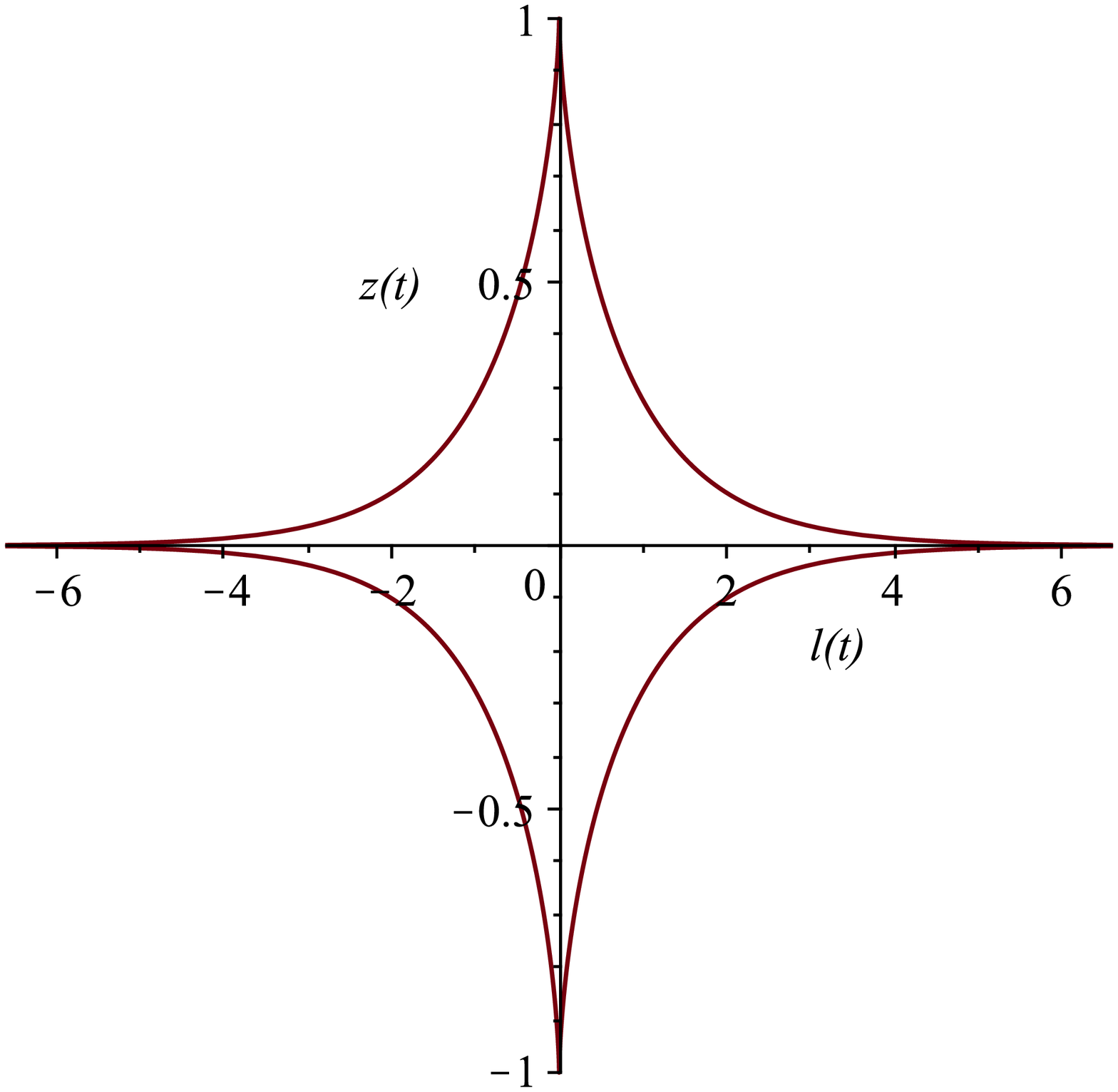}
		\caption{}\label{fig:pic5-1}
	\end{subfigure}  \qquad
	\begin{subfigure}[b]{0.40\textwidth}
		\centering
		\includegraphics[width=\textwidth]{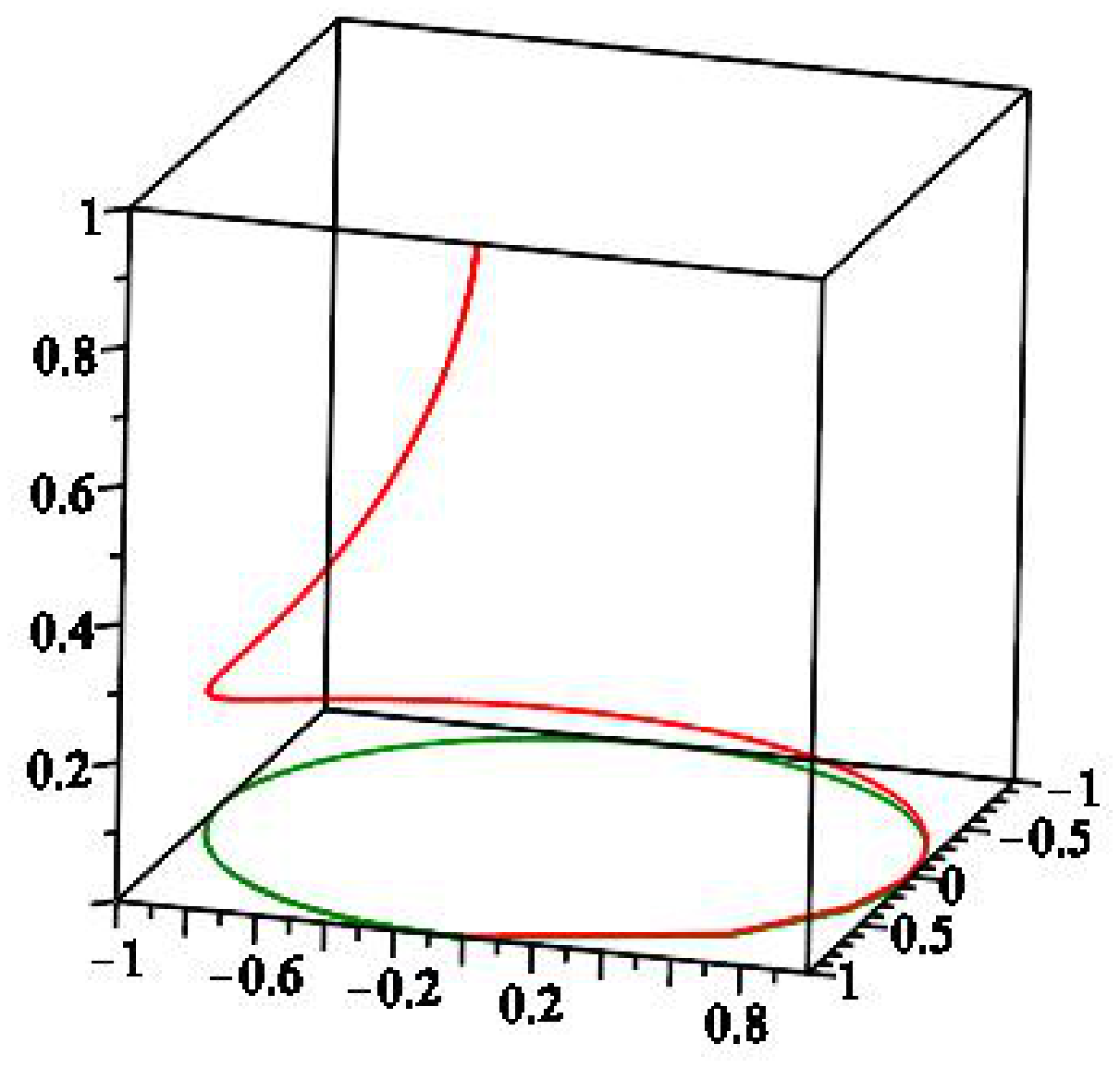}
		\caption{}\label{fig:pic5-2}
	\end{subfigure}
	\caption{}
\end{figure}

This relation is demonstrated in Figure \ref{fig:pic5-1}.
In Figure \ref{fig:pic5-2}, we show an example of lifting
of a unit circle,
given positive $z$ and $l$. In fact, the space curve
starts from the height of one unit above the circle and never
intersects with it.

\medskip
\textbf{6.} $h(a)= \ln{a} $ 

The relation between functions $l(\tau)$ and $z(\tau)$ is
\[
\sqrt{ e^{2z}-1}-\arctan{\sqrt{ e^{2z}-1}} = \pm  l(\tau).
\]

\begin{figure}[h]
	\centering
	\begin{subfigure}[b]{0.36\textwidth}
		\includegraphics[width=\textwidth]{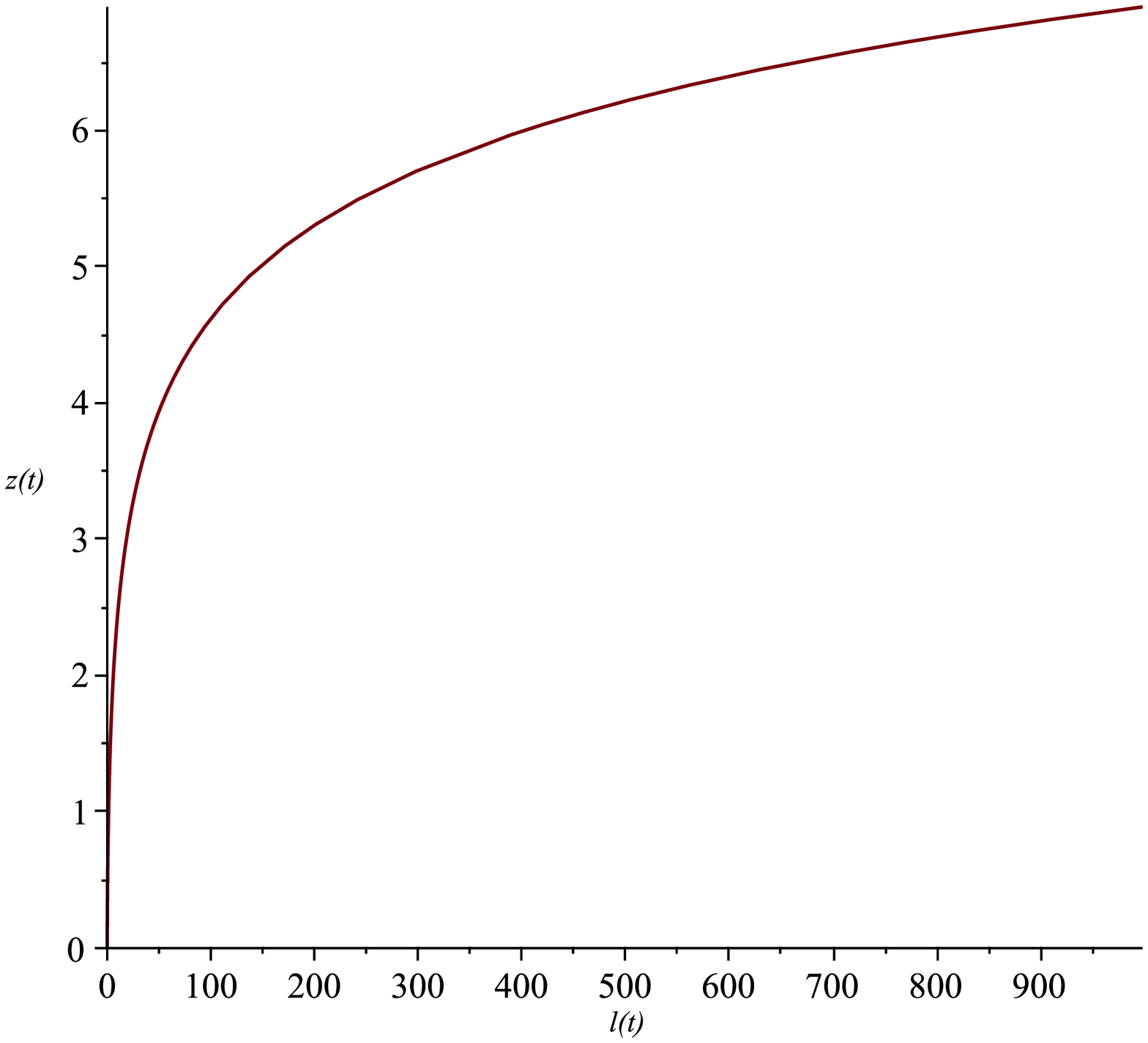}
		\caption{}\label{fig:pic6-1}
	\end{subfigure}  \qquad
	\begin{subfigure}[b]{0.40\textwidth}
		\centering
		\includegraphics[width=\textwidth]{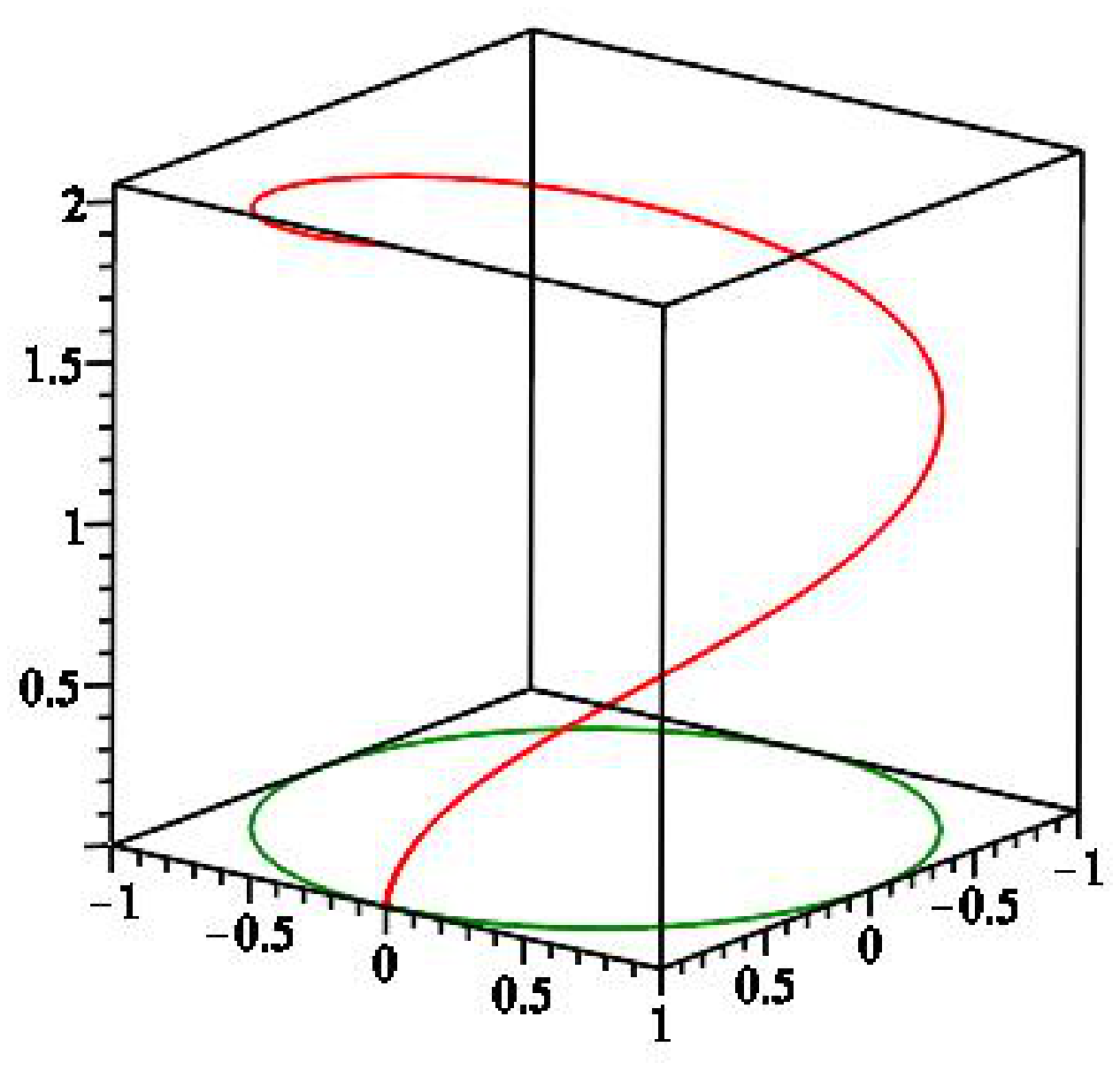}
		\caption{}\label{fig:pic6-2}
	\end{subfigure}
	\caption{}
\end{figure}

In Figure \ref{fig:pic6-1}, we show
this relation when $l$ is positive. In Figure \ref{fig:pic6-2}
the corresponding lifting of a unit circle is demonstrated.

\textbf{Acknowledgments.} The research was partially supported by RFBR Grant No 18-29-10013.


\end{document}